\documentclass[jgrga]{AGUTeX}

\usepackage[pdftex]{graphicx}
\setkeys{Gin}{draft=false}
\authorrunninghead{CLARKE ET AL.}

\titlerunninghead{JOVIAN DECAMETRIC EMISSION WITH LWA1}

\authoraddr{Corresponding author: T.~E.~Clarke,
Naval Research Laboratory, Code 7213, 4555 Overlook Ave SW, Washington, DC 20375, USA.
(tracy.clarke.ca@nrl.navy.mil)}

\begin{document}

%
%

\title{Probing Jovian Decametric Emission with the Long Wavelength Array Station 1}

%
%

\authors{T.~E.~Clarke\altaffilmark{1},
C.~A.~Higgins\altaffilmark{2}, Jinhie Skarda\altaffilmark{3}, Kazumasa Imai\altaffilmark{4},
Masafumi Imai\altaffilmark{5}, Francisco Reyes\altaffilmark{6},
Jim Thieman\altaffilmark{7},
Ted Jaeger\altaffilmark{8}, Henrique Schmitt\altaffilmark{1}
Nagini Paravastu Dalal\altaffilmark{9}, Jayce Dowell\altaffilmark{10}, 
S.~W.~Ellingson\altaffilmark{11}, Brian Hicks\altaffilmark{1}, Frank Schinzel\altaffilmark{10},
G.~B.~Taylor\altaffilmark{10}}

\altaffiltext{1}{Naval Research Laboratory, Code 7200, Washington, DC, USA.}

\altaffiltext{2}{Department of Physics and Astronomy, Middle Tennessee State University, 
Murfreesboro, TN, USA.}

\altaffiltext{3}{Department of Electrical Engineering, Stanford University, Stanford, CA, USA.}

\altaffiltext{4}{Department of Electrical Engineering and Information Science, Kochi National College of Technology, Kochi, Japan.}

\altaffiltext{5}{Department of Geophysics, Kyoto University, Japan}

\altaffiltext{6}{Department of Astronomy, University of Florida, Gainesville, FL, USA.}

\altaffiltext{7}{NASA/GSFC, Code 6901, Greenbelt, MD, USA.}

\altaffiltext{8}{Johns Hopkins University Applied Physics Laboratory, Laurel, MD, USA.}

\altaffiltext{9}{Work done as ASEEE at NRL, Code 7200, Washington, DC, USA.}

\altaffiltext{10}{Department of Physics and Astronomy, University of New Mexico, Albuquerque, NM 87131, USA.}

\altaffiltext{11}{Bradley Department of Electrical \& Computer Engineering, Virginia Tech, Blacksburg, VA 24060, USA.}

%
%

\begin{abstract}
New observations of Jupiter's decametric radio emissions have been
made with the Long Wavelength Array Station 1 (LWA1) which is capable
of making high quality observations as low as 11 MHz. Full Stokes
parameters were determined for bandwidths of 16 MHz. Here we present
the first LWA1 results for the study of six Io-related events at
temporal resolutions as fine as 0.25 ms. LWA1 data show excellent
spectral detail in Jovian DAM such as simultaneous left hand circular
(LHC) and right hand circular (RHC) polarized Io-related arcs and
source envelopes, modulation lane features, S-bursts structures,
narrow band N-events, and interactions between S-bursts and
N-events. The sensitivity of the LWA1 combined with the low radio
frequency interference environment allow us to trace the start of the
LHC Io-C source region to much earlier \mbox{CML\,{\sc iii}} than
typically found in the literature. We find the Io-C starts as
  early as \mbox{CML\,{\sc iii}} = 230$^\circ$ at frequencies near 11
  MHz. This early start of the Io-C emission may be valuable for
  refining models of the emission mechanism. We also detect
modulation lane structures that appear continuous across LHC and RHC
emissions, suggesting that both polarizations may originate from the
same hemisphere of Jupiter. We present a study of rare S-bursts
detected during an Io-D event and show drift rates are consistent with
those from other Io-related sources. Finally, S-N burst events are
seen in high spectral and temporal resolution and our data strongly
support the co-spatial origins of these events.
\end{abstract}

%
%

\begin{article}

%
%

\section{Introduction}

Strong bursts of decametric radio emission below 40 MHz were
identified by \citet{burke55} as originating from Jupiter. The
emission in the decametric range (referred to as {\sc DAM} range
hereafter) occurs from 3 MHz to 39.5 MHz, although specific
  Io-related {\sc DAM} arcs have been seen at frequencies below 2 MHz
  \citep{zarka01,queinnec98,alexander81}. Jovian {\sc DAM} has a peak of the
  spectral power at a frequency of about 10 MHz, and average fluxes of
  $\> 10$ MJy\footnote{1 Jy = $10^{-26}$ Watt m$^{-2}$ Hz$^{-1}$}, as
  well as S-bursts that can be 100 times stronger \citep{zarka98}. The
  lower-limit on observations from ground-based observatories is
  generally above 5 to 10 MHz due to attenuation by the Earth's
  ionosphere.

From ground-based and space-based observations, it has been long
  known that much of the Jupiter {\sc DAM} emission is influenced by the
  satellite Io \citep{bigg64}, and that some emissions not related to Io
  have been correlated with solar wind parameters
  \citep{barrow86,hess12}. For a review see \citet{carr83}. These
coherent emissions are thought to be the result of instabilities near
the local electron cyclotron frequency, called the cyclotron maser
instability (CMI) mechanism \citep{treumann06, zarka98, wu79}. The CMI emission
is beamed in a hollow cone with thickness of a few degrees
\citep{kaiser2000} which has a large half-angle of
60$^\circ$ to 90$^\circ$ relative to the local magnetic field
direction \citep{hess08, queinnec98, dulk67}. The {\sc DAM} emission is observed at Earth
when the edge of the emission cone sweeps over the observer. General
properties of the {\sc DAM} emission include 1) high brightness
temperatures, $> 10^{15}$ K, 2) emission at frequencies near the
X-mode cutoff which is close to the local electron cyclotron frequency
from both northern and southern hemisphere emission regions, and 3)
100\% circular or elliptical polarization \citep{clarke04}.

The Io-controlled Jovian emission regions are designated Io-A and Io-B
for the primarily right hand circular (RHC) emission sources in the
northern hemisphere while the predominantly left hand circular (LHC)
sources in the southern hemisphere are designated Io-C and Io-D. The handedness of the polarized emission is set by the direction of
  propagation of the emission relative to the local magnetic field
  direction as seen by the observer \citep{zarka88, thieman79}. These
Io-related emission regions are defined by their occurrence
probability in the central meridian longitude (\mbox{CML\,{\sc iii}})
and Io-phase ($\phi_{Io}$) plane \citep{carr83}.

\begin{table*}[tb!]
\caption{LWA1 Array Comparison with other Decameter Telescopes used for Jovian Study}
\centering
\begin{tabular}{cccccc}
\hline
Name & Dipole Type & Number of Dipoles & Freq & Sensitivity\tablenotemark{1}, BW & Reference\\
 & Type & x Polarizations & (MHz) & mJy, MHz & \\
\hline
LWA1 & Bowtie & 256 x 2 & 10-88 & 17, 16 & 1\\
Nan\c{c}ay Decameter Array& Conical Log Spiral & 72 x 2 & 10-120 & 1670, 1 & 2\\
UTR-2 & Cylinder & (1440 + 600) x 1 & 8-30 & 10000, 3 & 3\\
LOFAR LBA Station & Wire & 48 x 2 & 10-90 & 40, 16 & 4\\
\hline
\end{tabular}
\tablenotetext{1}{Zenith sensitivity at 20 MHz in 1 hr at listed bandwidth (BW). Values from \citet{taylor12} except NDA from \citet{lecacheux91}.}
\tablecomments{{\bf References:} 1. \citet{ellingson13a}, 2. \citet{boischot80}, 3. \citet{braude78}, 4. \citet{vanHaarlem}}
\label{tbl:ant}
\end{table*}

There are many types of fine structures seen in Jovian decametric
radio emission. Most of these structures have been seen by other
ground-based and spacecraft observers \citep{zarka98, carr83}. Common features include emission frequency envelopes that differ for
  Io and non-Io related events, as well as spectral arc structures
  that depend both on the frequency and the Io/non-Io source
  parameters \citep{hess08, carr83}.  Jupiter DAM emission can be
further characterized as consisting of a set of long bursts (L-bursts)
which last several minutes but have temporal substructure (few
seconds) modulated by interplanetary scintillations, short
milli-second bursts (S-bursts), as well as narrow band emission
(N-events). While the L-bursts are modulated by scintillation, the
S-bursts show modulation and both simple and complex structure that is
intrinsic to the source regions \citep{arkypov09, hess07, carr99,
  carr83}.  The N-events have been observed since the 1960s and show
widely varying bandwidths from 15 kHz to 200 kHz and have durations
from seconds to many minutes \citep{riihimaa85, carr83}. The N-events
also show complex structure such as splitting and interactions with
S-bursts, called S-N events, \citep{oya02, riihimaa81, boischot80} and
are seen both with smooth and erratic frequency drift rates.

The basic envelope and arc structure of bursts comes from the rotating
conical beam pattern and Alfv\'{e}n wave propagation and reflection along Io's flux tube
\citep{hess08, queinnec98}. L-bursts are from CMI emission occurring as trains
of emission interrupted by modulations in Jupiter's Io plasma torus as
well as interplanetary and terrestrial ionospheric scintillations
\citep{carr83}. S-bursts are CMI emission thought to be produced by 5 keV
electrons accelerated along magnetic field lines connecting to Io
\citep{hess07, Zarka96}. S-N event emissions are not well understood but they
are believed to be interactions at the source between precipitating
electrons and resonant emission regions \citep{oya02, carr83}.

There are also additional structures seen within the Jovian emission
such as the modulation lanes first seen by \citet {Riihimaa68}. These
lanes are seen as groupings of equally-spaced intensity variations
running across multiple L-bursts. The modulation lanes are known to
slope upward or downward in frequency and the typical separation of
the lanes is of order 2 s. The modulations lanes can generally be
traced over large frequency ranges of 5 to 10 MHz. The origin of these
modulation lanes is not completely understood but one model, which
fits well with the data, demonstrates that the lanes originate from an
interference pattern generated by passage of the Jovian decametric
emission through vertical plasma density fluctuations along magnetic
field lines located near Io's orbit close to the longitude of the
sub-Earth point \citep{imai92}. This model is capable of yielding new
and detailed information regarding the Jovian decametric sources and
the environment near Io's orbit in which the interference-producing
plasma structure is located.

Despite extensive study of Jovian decametric emission since its
discovery in 1955, there are still many open issues such as the nature
of the emission and particle acceleration mechanism(s), the three
dimensional structure and location of the DAM emission cones, the
local plasma densities in the satellite flux tubes, processes
influenced by the solar wind and by Io, and the physical processes
responsible for the fine spectral signatures (modulation lanes,
S-bursts, N-events, etc.). In this paper we present an overview of our
Jovian observing campaign using the newly developed Long Wavelength
Array Station One (LWA1). These observations agree well with
  studies from the literature, and allow us to extend our
  understanding of the Jovian emission by showing the early start of
  Io-C emission at low frequencies (Section~\ref{ssect:AC}) as well as
  revealing the structure of rare S-bursts seen during Io-D events
  (Section~\ref{sect:D_FS}). These results show the power of the LWA1
  instrument for Jovian research and may lead to refinements in our
  understanding of the Jovian emission processes.

\section{Instrument: The Long Wavelength Array Station One}

Observations of Jovian decametric emission were obtained using the
first station of the Long Wavelength Array. LWA1 is a
low frequency radio dipole array located in New Mexico
near the National Radio Astronomy Observatory's Jansky Very Large
Array. We summarize relevant instrumental parameters below and refer
the reader to \citet{ellingson13a} for a more complete description of the
instrument.

LWA1 consists of 256 dual polarization antennas.  Each antenna
  has an active-balun-preamp-filter design that gives 6 dB gain above
  the Galactic background over the bandwidth 20-80 MHz without
  increasing the overall noise of the system. The tuneable range of
  LWA1 is 10-88 MHz. Details of the ``active'' antenna system are
  presented in \citet{hicks12}.

The outputs of the LWA1 antennas are digitized, delayed, and summed to
form up to four independently pointed beams on the sky. Each beam has
two fully independent frequency tunings with bandwidths that can be
selected between roughly 0.2 MHz and 16 MHz.  Each LWA1 beam provides
dual orthogonal linear polarizations such that it is possible to
reconstruct the full Stokes parameters for each tuning. LWA1 also
operates with a transient buffer mode that digitizes each polarization
from each antenna to provide all-sky imaging \citep{ellingson13a}.

The system equivalent flux density (SEFD) of an array provides a
measure of the combined sensitivity of the array and receiver
system. SEFD can be thought of as the flux density of an unresolved
radio source in the beam that doubles the measured system power
relative to measurements without the source. The SEFD of the LWA1 is
measured by \citet{ellingson13a} as 16.1 kJy (single polarization) at
74.03 MHz using Cygnus A which is in a relatively bright part of the
sky. Table II of \citet{ellingson13a} reports the single polarization
LWA1 SEFD for Cygnus A at 37.9, 28.8, and 20.5 MHz to be 21.5, 18.2,
and 47.0 kJy. The SEFD measurement will decrease away from the
Galactic plane and also depends inversely on elevation. The impact of
source location and elevation can be seen in Figure 12 of the
\citet{ellingson13a} paper where measurements at several frequencies
and zenith angles toward a variety of sources in and out of the
Galactic plane are presented.  The SEFD of the LOFAR low-band antennas
(LBA) as shown in \citet{vanHaarlem} at roughly 38 MHz is 32 kJy for
the LBA outer core stations and 38 kJy for the LBA Inner core
stations. At 30 MHz, the lowest measurements shown in
\citet{vanHaarlem}, the LOFAR SEFD is roughly 38 kJy for both LBA
inner and outer core stations.  We note that comparable SEFD
measurements are not available for the Nan\c{c}ay Decameter Array (NDA
hereafter) and UTR-2 so it is not possible to compare them on a
consistent (i.e.\ SEFD) basis with the LWA1. \citet{taylor12} compare
the zenith sensitivity of low frequency instruments (except NDA) for 1
hour observations at typical bandwidths for the instruments in their
Figure 2. In Table~\ref{tbl:ant} we provide a comparison of dipole
type, number of dipoles, number of polarizations, frequency coverage
and Zenith sensitivity of several low frequency instruments typically
used for Jovian decametric research. We use the numbers from Figure 2
of \citet{taylor12} and add the sensitivity of NDA from
\citet{lecacheux91}.

\begin{figure*}[tb!]
\noindent\includegraphics[width=3.6in]{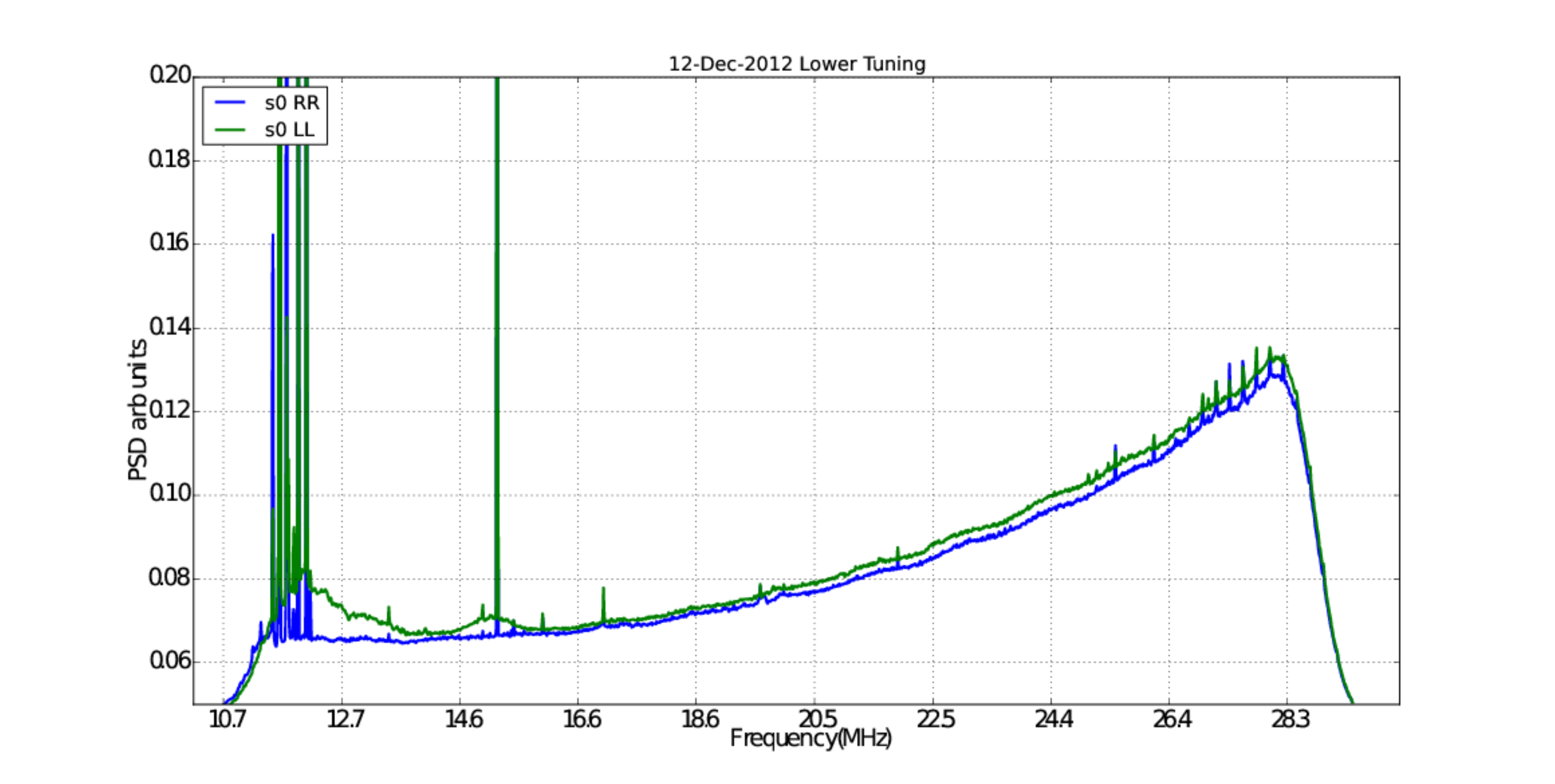}
\noindent\includegraphics[width=3.6in]{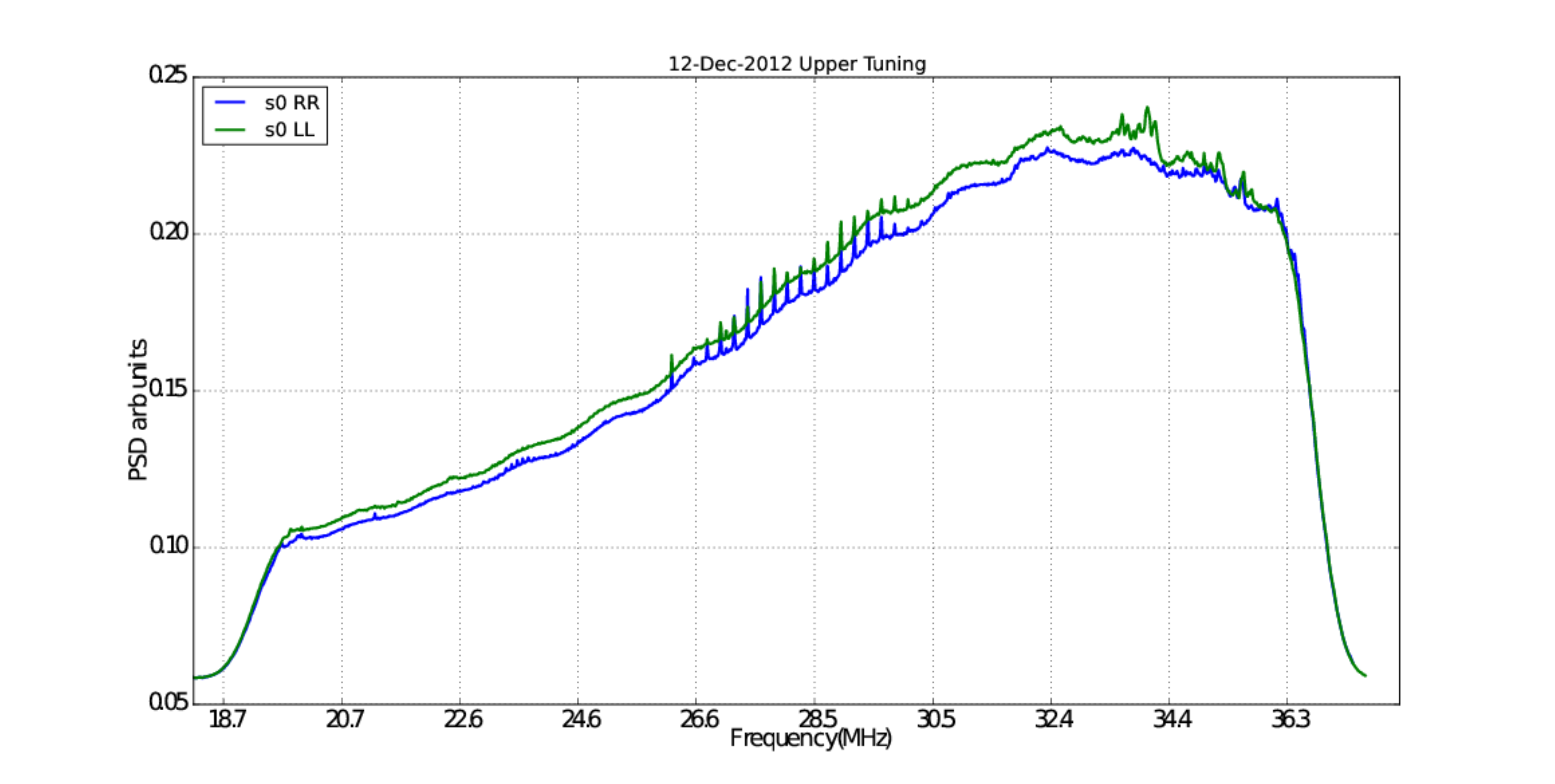}
\caption{Top panel shows the LWA1 bandpass of the Jupiter beam for R
  and L polarizations (blue and green, respectively) for the lower
  tuning ($\nu_{center}$=20 MHz). The bandpasses were made using 120 s
  of data during a non-bursting portion of the Jovian observing. There
  has been no removal of RFI. Bottom panel shows the bandpass at the
  same time for the upper tuning ($\nu_{center}$=28 MHz).}
\label{fig:BP}
\end{figure*}

\subsection{Jovian Program Observing Setup}

Jovian decametric emission is confined to the lower half of the LWA1
frequency band below 40 MHz. Observation campaigns were undertaken
mainly within the typical Jovian observing season when Jupiter is
visible at night, although a few observations were also taken outside
this season during daytime. The observing schedule was determined
using the known Io-related burst probability
distribution\footnote{Radio Jupiter Pro software based on U. Florida
  probability tables}. We mainly targeted long duration bursts
(greater than 30 minutes) at high elevations where the burst
probability passed through the highest probability regions. 

The observations were all taken in wide bandwidth mode with a total
observing bandwidth of 16 MHz. The majority of the observations were
obtained with tunings centered at 20 and 28 MHz. These passbands contain
significant overlap allowing us to cover the entire frequency range of
10.2 MHz to 37.8 MHz with the frequency range of 12 - 36 MHz having
nominal sensitivity. Several of the Io-B events were observed at
higher tunings of 22 and 30 MHz due to the generally higher peak frequency
emission observed from these events, thus allowing nominal sensitivity
over the 14 - 38 MHz range. In addition to the beam tracking Jupiter,
we also used identical tunings for one to two off-beams which are used
for data quality and instrumental bandpass analysis. These off-beams
were centered on a range of targets including several well-established
radio calibrators such as the supernova remnants Cas-A, and the Crab
Nebula, as well as the extragalactic radio galaxy Cyg-A. Additional
off-beam data was also obtained from the north celestial pole.

The LWA1 beam full width at half maximum near zenith ranges from
8.1$^\circ$ at 20 MHz to 5.8$^\circ$ at 30 MHz
\citep{ellingson13a}. Although most bursts were targeted for high
elevations, some bursts were tracked for very long times and reached
low elevations of roughly 10$^\circ$. At these elevations the LWA1
beam will be significantly elongated.

In this paper we present a subset of the LWA1 Jovian observations to
provide on overview of the capabilities of the system. We will present
more detailed analysis of the individual events in future papers.

\section{Data Analysis}

LWA1 data are recorded as raw voltage streams without
channelization. These data were processed first through the LWA
Software Library tools \citep{jayce12} using custom built Python
scripts. Data were read from the LWA1 binary data format and processed
through a Fourier transform spectrometer to obtain frequency-time
spectrograms of the observation. Data are recorded in X and Y
polarizations but are converted to the desired polarization basis of
[X,Y], [R,L], or Stokes [I,Q,U,V] prior to constructing the
spectrogram of the data. Frequency and time channelization are
flexible and the standard configuration used for the Jovian
observations allows for time bins as fine as 0.21 ms with associated
frequency channelization as fine as 5 kHz.

All bursts were initially run averaging the full observing session
data to a 2 second time interval with a spectral resolution of 10 kHz
to provide a coarse time sampling of the entire burst structure. The
goal of this process was to obtain an overview of the burst in Stokes
I and V to examine the full burst structure and identify interesting
regions for study. We present full-burst spectrograms of several
events in Section~\ref{sect:full_bursts}. Once regions of Jupiter
  bursts were identified in the spectrum at 2 s resolution, higher
  time resolution power spectra were formed with resolutions of up to
  0.25 ms. Details of the fine structure of the bursts are presented
  in Section~\ref{sect:fine_structure}.

Due to the wide instantaneous observing bandwidth of the LWA1, the
instrument response will change across the observing bandwidth. Using
our off-beams as well as the non-bursting sections of the Jovian beam
we examined the shape and stability of the bandpass. In
Figure~\ref{fig:BP} we show the median bandpass for the R and L
polarizations for both the upper and lower tunings of an Io-A/Io-C
burst observed 01-Dec-2012.  These plots were made without any
excision of radio frequency interference (RFI) using non-bursting
times of the Jovian beam. We see a slight offset in the power of the R
and L polarizations which is due to a known offset in the X and Y
dipole responses \citep{Dowell11, Ellingson2010}. We have
calibrated the offset between the R and L polarizations using both the
Jovian beam as well as a simultaneous off-beam on Tau A for several of
our Jovian observing sessions.  We find that the L polarization is
consistently higher than the R polarization. The percentage difference
between the two beams shows variations with time but remains below 5\%
for observations presented herein. This is reasonable since the
observations were taken at the same frequency and at similar azimuth and zenith angles.

All data presented in this paper have been converted to Stokes V
following the convention where positive Stokes V is RHC. We note that
we cannot correct the bandpass of the individual polarizations during the
bursting periods thus the Stokes V results will have a negative bias
(due to the different R and L bandpass amplitudes) of up to 5\% in the
spectrogram plots. This offset is minor and does not impact the
handedness of any emission presented in this paper. Furthermore,
  the circular polarization of Jupiter {\sc DAM} emission events is shown
  to have little variation over long durations \citep{dulk94}.

\begin{table*}[tbh!]
\caption{Jovian Burst Details}
\centering
\begin{tabular}{ccccccc}
\hline
Date & MJD & Tunings & Io-type & Polz.\tablenotemark{1} & Io phase range &  \mbox{CML\,{\sc iii}} range\\
 & & (MHz) & & & ($^\circ$) & ($^\circ$)\\
\hline
10-Mar-2012 & 055996 & 26 & Io-C & LHC/RHC & 238.6 - 249.8 & 302.0 - 349.0 \\
11-Mar-2012 & 055997 & 26 & Io-B/Io-D & RHC/LHC & 82.2 - 105.8 & 93.6 - 194.6\\
24-Sep-2012 & 056194 & 20, 28 & Io-B & RHC & 71.4 - 92.6 & 119.8 - 210.4\\
01-Dec-2012 & 056262 & 20, 28 & Io-A/Io-C & RHC/LHC & 210.7 - 253.1 & 184.2 - 5.6 \\
27-Dec-2012 & 056288 & 22, 30 & Io-B/Io-D & RHC/LHC & 89.9 - 106.8 & 77.4 - 150.0\\
06-Mar-2013 & 056357 & 28 & Io-A/Io-C & LHC/RHC & 227.0 - 253.3 & 249.0 - 0.0 \\
\hline
\end{tabular}
\tablenotetext{1}{Handedness of polarization observed in burst where first polarization listed is
the dominant polarization.}
\label{tbl:obs}
\end{table*}

\begin{table*}[bh!]
\caption{Observing Session Details}
\centering
\begin{tabular}{cccccc}
\hline
Date & Start Time & Duration & Sunrise & Sunset & Elevation\\
 & (UT) & (hours) & (MST\tablenotemark{1}) & (MST) & ($^\circ$)\\
\hline
10-Mar-2012\tablenotemark{2} & 23:00:41 & 1.3 & 06:27 & 18:15 & 66 - 54 \\
11-Mar-2012\tablenotemark{2} & 23:03:01 & 2.8 & 06:26 & 18:16 & 65 - 35 \\
24-Sep-2012\tablenotemark{2} & 11:00:00 & 2.5 & 06:00 & 18:04 & 74 - 63 \\
01-Dec-2012 & 08:15:00 & 5.0 & 06:58 & 17:01 & 68 - 7 \\
27-Dec-2012 & 06:30:00 & 2.0 & 07:15 & 17:09 & 66 - 42 \\
06-Mar-2013\tablenotemark{2} & 22:45:00 & 3.1 & 06:32 & 18:11 & 64 - 68 \\
\hline
\end{tabular}
\tablenotetext{1}{MST zone (Mountain Standard Time) is UT-7 hours.}
\tablenotetext{2}{Observing session includes at least some daytime data.}
\label{tbl:obs_time}
\end{table*}

The LWA1 has not yet been characterized for polarization response thus
observations at large zenith angles are expected to have significant
polarization leakage across the Stokes parameters. Determining the
polarized response of a dipole array is considerably more difficult
than for a single dish telescope.  The reasons for this are: (1)
dipole arrays produce beams with higher side lobe levels, therefore a
significant contribution to the beam power comes from beyond the
primary beam; (2) the dipole array beam shape and sensitivity vary
significantly across the sky.  The first factor violates the usual
assumption that the source polarization properties are constant in
time (or change only as a function of parallactic angle), and the
second factor violates the assumption that the instrumental
polarization is constant in time.

We have minimized the impact of the lack of polarization calibration
by observing Jovian bursts at high elevations near transit. To
characterize the impact of this leakage on the measured polarization,
we have compared the fractional circular polarization measured for
Jovian bursts with measurements from the literature. Using Io-B and
Io-C bursts we have measured the fractional circular polarization of
the emission over zenith angles from $\sim$ 20$^\circ$ to over
50$^\circ$ and compared those measurements to statistical measurements
of fractional polarization of the same type of bursts observed with
the NDA as presented in \citet{dulk94}. We
find very good agreement between the LWA1 measured fractional
polarization and the NDA statistical results up to zenith
angles of 40$^\circ$. At larger zenith angles we see an elevation
dependent fractional polarization which is due to the lack of
polarization calibration. A similar effect is seen observing Cygnus A
over a variety of zenith angles where the fractional circular
polarization shows a slow increase with increasing zenith angle beyond
roughly 50$^\circ$ (J.\ Hartman, 2012, private communication). We would not
expect that the lack of polarization calibration would change the
handedness of the emission observed and, in fact, we find that the
handedness of the emission observed in all bursts recorded with the
LWA1 matches that expected from the literature. Due to the lack of
polarization calibration, we do not measure polarization fraction
within this paper and report mainly the handedness of the emission as
well as source structure.

RFI is present in the data in both persistent form with known emitters
spread throughout the spectrum as well as intermittent RFI events \citep{obenberger11}.
Observations were preferentially scheduled at night when RFI levels
are lower but many of the observations had to be taken at least
partially during the daytime to target high probability events in the
observing cycle.  For the majority of the Jovian observations, we found
that the RFI only impacted a small amount of the spectrum and thus we
have not excised these regions. During quiescent RFI times we were
easily able to study Jovian emission down to frequencies of 12 MHz or
lower (see for example Figure~\ref{fig:262}).

\section{Morphology and Polarization properties of Several Io-Related Decametric Bursts}
\label{sect:full_bursts}

LWA1 observations of Jupiter were targeted during times of
Io-controlled emission. In this section we present analysis of bursts
from all four of the Io-controlled emissions regions. In
Table~\ref{tbl:obs} we list event details of each burst discussed
in this paper while the observing session details are given in
Table~\ref{tbl:obs_time}. In Figure ~\ref{fig:prob} we show a
probability plot of the Io phase vs \mbox{CML\,{\sc iii}} with
the positions of Jupiter marked for each of the bursts discussed in
this paper.

\subsection{Io-A/Io-C}
\label{ssect:AC}

\begin{figure*}[bh!]
\centering
\noindent\includegraphics[width=40pc]{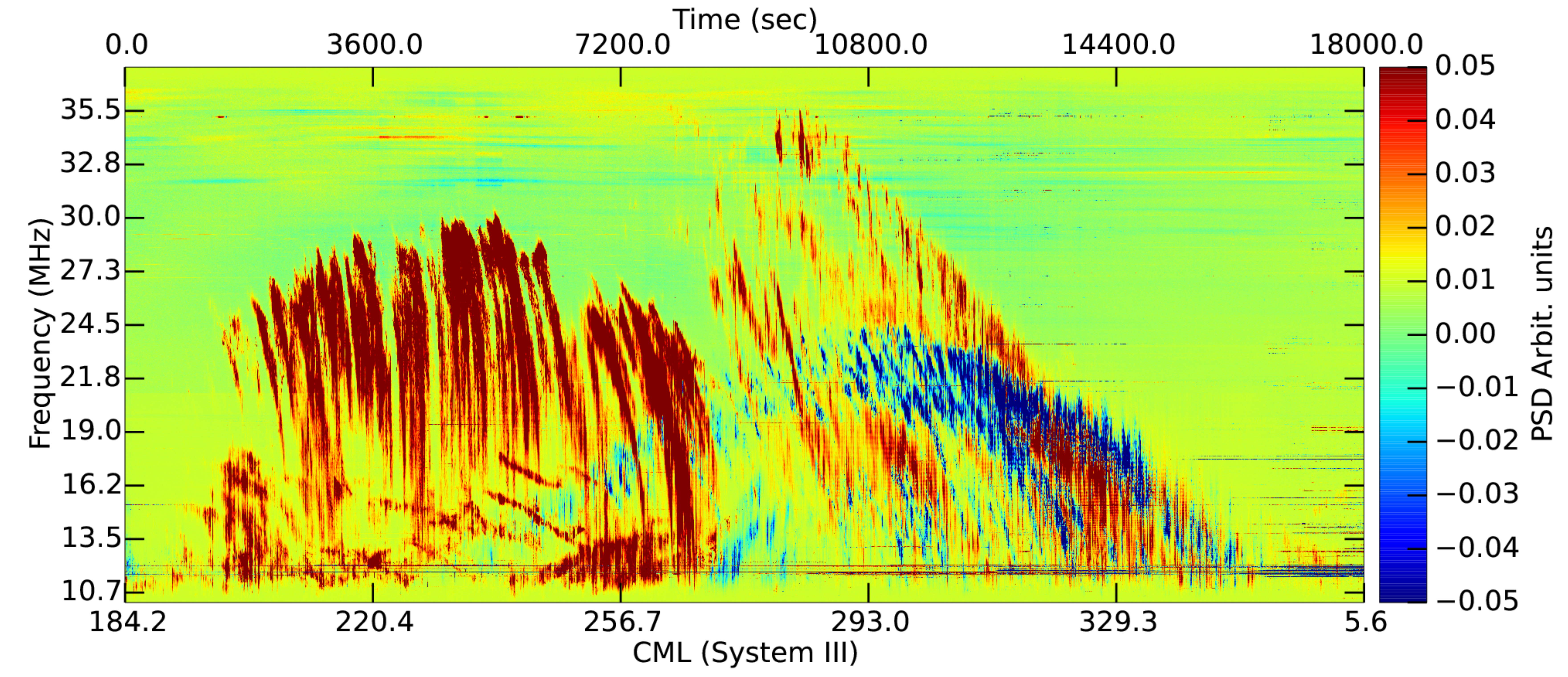}
\caption{Io-A/Io-C burst recorded 01-Dec-2012 for a total duration of
  5 hours. Both the upper and lower tunings have been merged to show
  the entire frequency range covered along the Y axis. The elapsed
  time since the start of the observation is shown in the upper axis
  while the \mbox{CML\,{\sc iii}} range in shown on the lower 
  axis. The color scale is the power spectral density in arbitrary
  units shown in the right color bar. On this scale right hand circular emission is
  in red while left hand circular is in blue.}
\label{fig:262}
\end{figure*}

We show in Figure~\ref{fig:262} a Stokes V spectrogram of an Io-A/Io-C
event from 01-Dec-2012. The full observation lasts 5 hours and we have
combined the data from the two frequency tunings. The colorbar at the
right shows intensity (labeled as power spectral density or PSD) with
negative Stokes V (LHC) emission in blue and positive Stokes V (RHC)
emission in red on a scale that is linear in power in arbitrary
units. The ordinate axis shows observing frequency while the top abscissa
is labeled in time since the start of the observation
and the bottom abscissa is labeled in central meridian longitude system
III (\mbox{CML\,{\sc iii}}).

The Io-A event (\mbox{CML\,{\sc iii}} 185$^\circ$ - 270$^\circ$)
  shows classic vertex late arc structures, which are possibly caused
  by multiple Alfv\'{e}n wave reflections between the Io plasma torus
  and the source regions \citep{queinnec98, gurnett81}. We see
  additional (fainter) RHC emission between \mbox{CML\,{\sc iii}} =
  270$^\circ$ to roughly 350$^\circ$ where it appears to drop below
  our observing frequency. This second RHC emission component appears
  distinct from the main Io-A emission; the overall intensity is
  weaker and the envelope reaches a higher frequency of $\sim$35
  MHz. \citet{boudjada95} studied similar Io-C morphology using
  NDA data from 1986 to 1991 and identified
  one RHC component (called a great arc) and three LHC components
  (Type I, II, and III). They interpret their data as components of
  Io-C and suggest that both RHC and LHC emissions are coming from the
  same source in the same hemisphere. Recent modulation lane analysis
  (Section~\ref{sect:modulation}) also favors this interpretation;
  however, the different spectral structure and different frequency
  maxima of the RHC and LHC data are difficult to explain. We also
  note that bimodal polarizations emitted by the same source are not
  favored by theory \citep{melrose84}.

The LHC event begins at about \mbox{CML\,{\sc iii}} 230$^\circ$ and
increases in frequency to reach a maximum frequency of $\sim$24.5 MHz
before decreasing again in frequency to fall below our observing band
around \mbox{CML\,{\sc iii}}=350$^\circ$. We interpret this event as
an Io-C burst but we note that it begins well before the generally
accepted beginning value of \mbox{CML\,{\sc iii}}=280$^\circ$ for Io-C
events \citep{carr83}. Although the previously published source
boundaries are based on long-term statistics and are frequency
  dependent, individual Io-C events beginning at values prior to
\mbox{CML\,{\sc iii}} 280$^\circ$ have been shown before
\citep[e.g.][]{ray08, boischot81}. The CML boundaries were not
previous discussed, however, these new LWA1 observations suggest that
the \mbox{CML\,{\sc iii}} range for the Io-C source needs
refining. More precise source boundaries will be of value to theorists
and modelers to better know the active CML regions and their
associated magnetic field lines. Furthermore we hope the LWA1 data can
help disentangle the Io-A from the Io-C source(s) and be used to
determine the hemispheric source(s) for the Io-A and/or Io-C
components.

The vertex late arc structures characteristic of Io-C events are
clearly seen by the LWA1. In Figure~\ref{fig:262_L_zoom} we show a
higher resolution section of the LHC polarization in the lower
frequency tuning for the portion of the burst from 10800 s onward. We
can clearly see that the Io-C event splits into at least two different
structures near the end of the event. The very subtle, nearly
horizontal amplitude modulations seen near 19 MHz in the burst
emission are ionospheric Faraday lanes induced in the highly
elliptically polarized emission as it passes through the Earth's
ionosphere (see Figure~\ref{fig:288_FL} for more obvious Faraday lanes).

\begin{figure}[tbh!]
\noindent\includegraphics[width=3.3in]{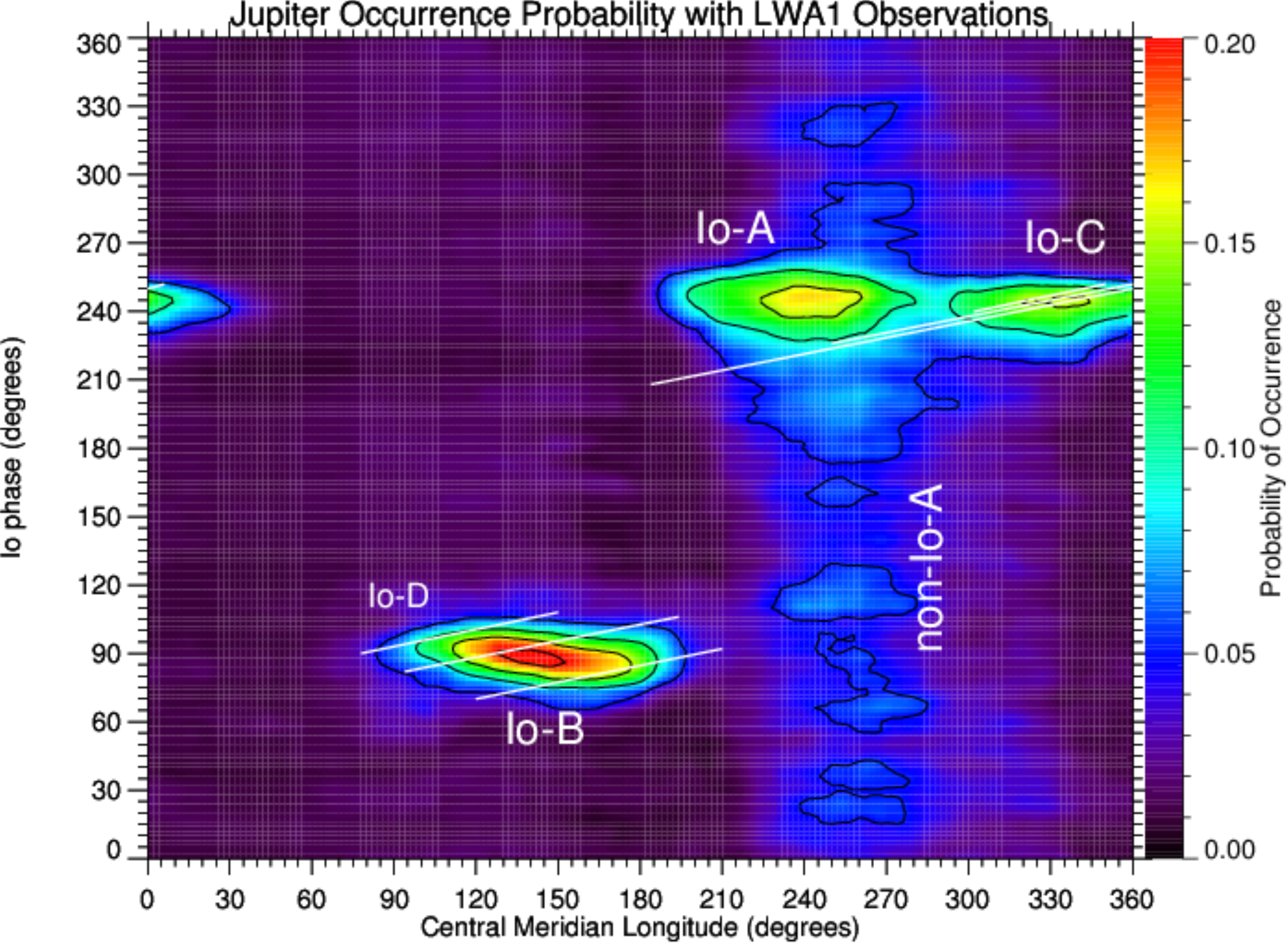}
\caption{Occurrence probability plot showing the Io phase vs
  \mbox{CML\,{\sc iii}}. The probability of occurrence is shown in the
    color bar where the probability was determined from the work of
    \citet{carr83}. The white sloped lines show the regions of the Io
    phase vs \mbox{CML\,{\sc iii}} space probed by the Jovian bursts
      presented in this paper (see Table~\ref{tbl:obs}).}
\label{fig:prob}
\end{figure}

\begin{figure*}[tb!]
\centering
\noindent\includegraphics[width=40pc]{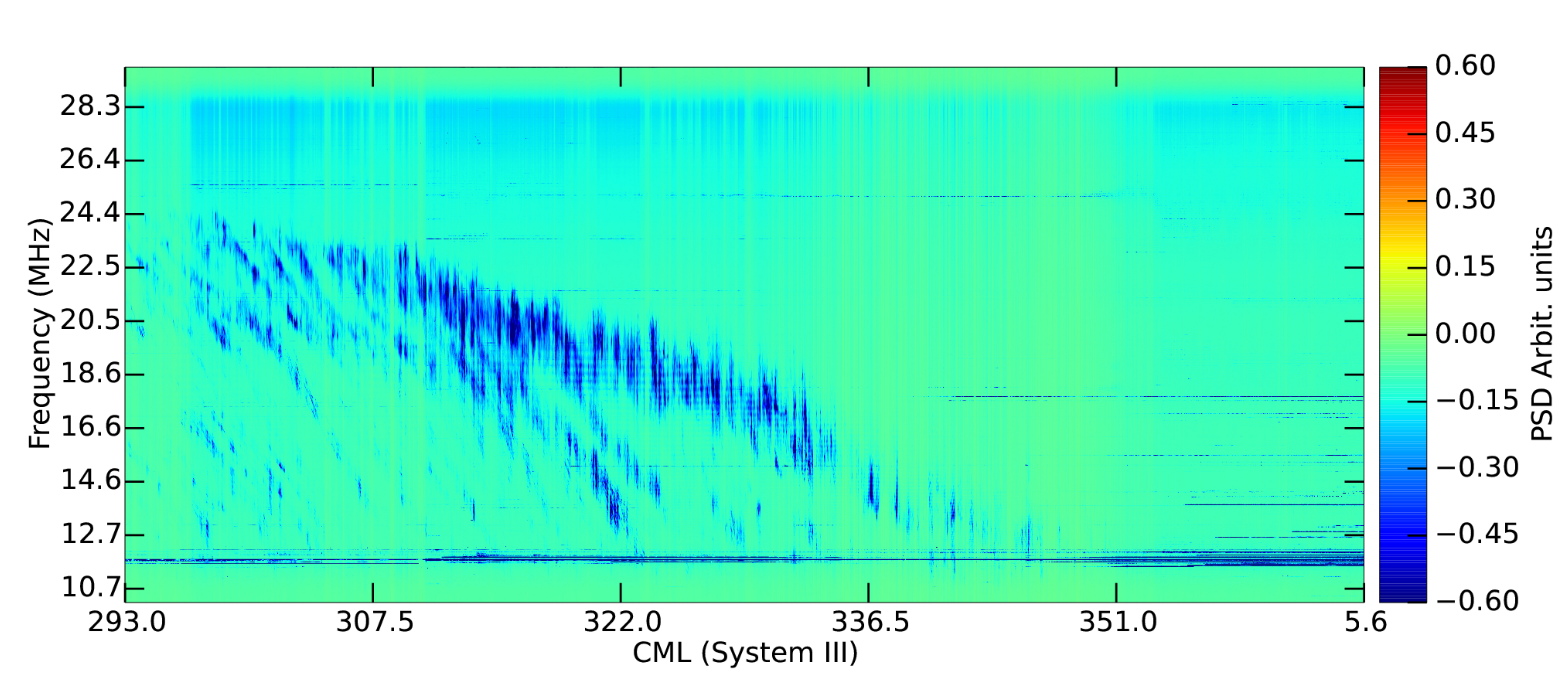}
\caption{Segment of the Io-C burst seen in Figure~\ref{fig:262} shown
  in higher resolution from CML\,{\sc iii}=293.0$^\circ$ to the end of
  the burst. The LHC polarization clearly splits into at least two different structures. 
}
\label{fig:262_L_zoom}
\end{figure*}

\begin{figure*}[bh!]
\centering
\noindent\includegraphics[width=40pc]{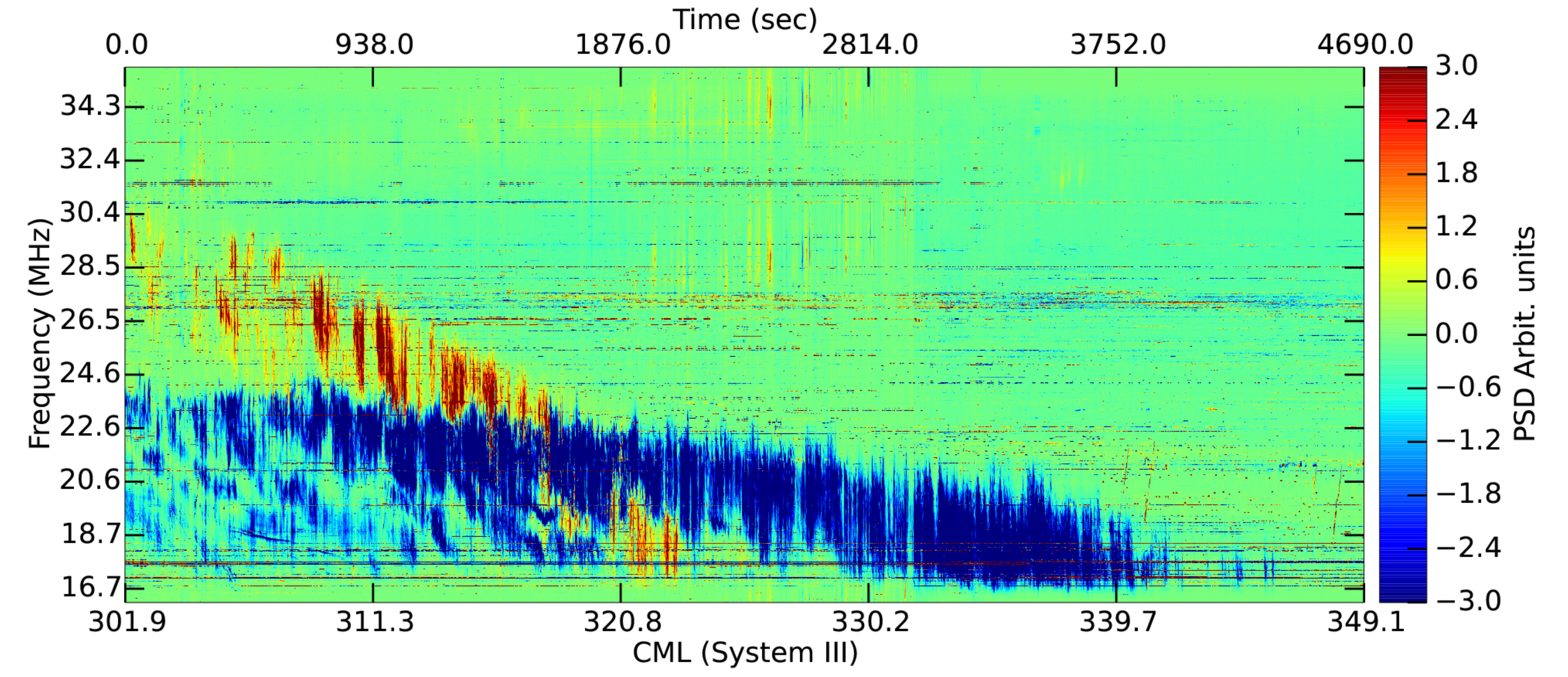}
\caption{Upper tuning (centered on 26 MHz) Stokes V spectrogram of an Io-C burst
  recorded 10-Mar-2012 for a total duration of 1.3 hours. 
}
\label{fig:996}
\end{figure*}

We have observed two other Io-C events with the LWA1 and find
remarkable similarity in the events. In Figure~\ref{fig:996} we show
the upper frequency tuning (centered on 26 MHz) for an Io-C event from
10-Mar-2012. RHC emission is seen from the observation start at
\mbox{CML\,{\sc iii}} = 302$^\circ$ until \mbox{CML\,{\sc iii}} =
323$^\circ$, where the emission drops below an observing frequency of
17 MHz. The LHC emission is seen from the observation start until it
falls below our band at $\sim$ \mbox{CML\,{\sc iii}} =
342$^\circ$. Similar to the burst described above, we interpret the
RHC emission as a 'Great arc' and the LHC emission as Io-C, most
likely coming from the southern hemisphere. The Io-C emission reaches
a maximum frequency of roughly 25 MHz and shows the same vertex late
arc structures as the previous burst.

\begin{figure*}[h!]
\centering
\noindent\includegraphics[width=40pc]{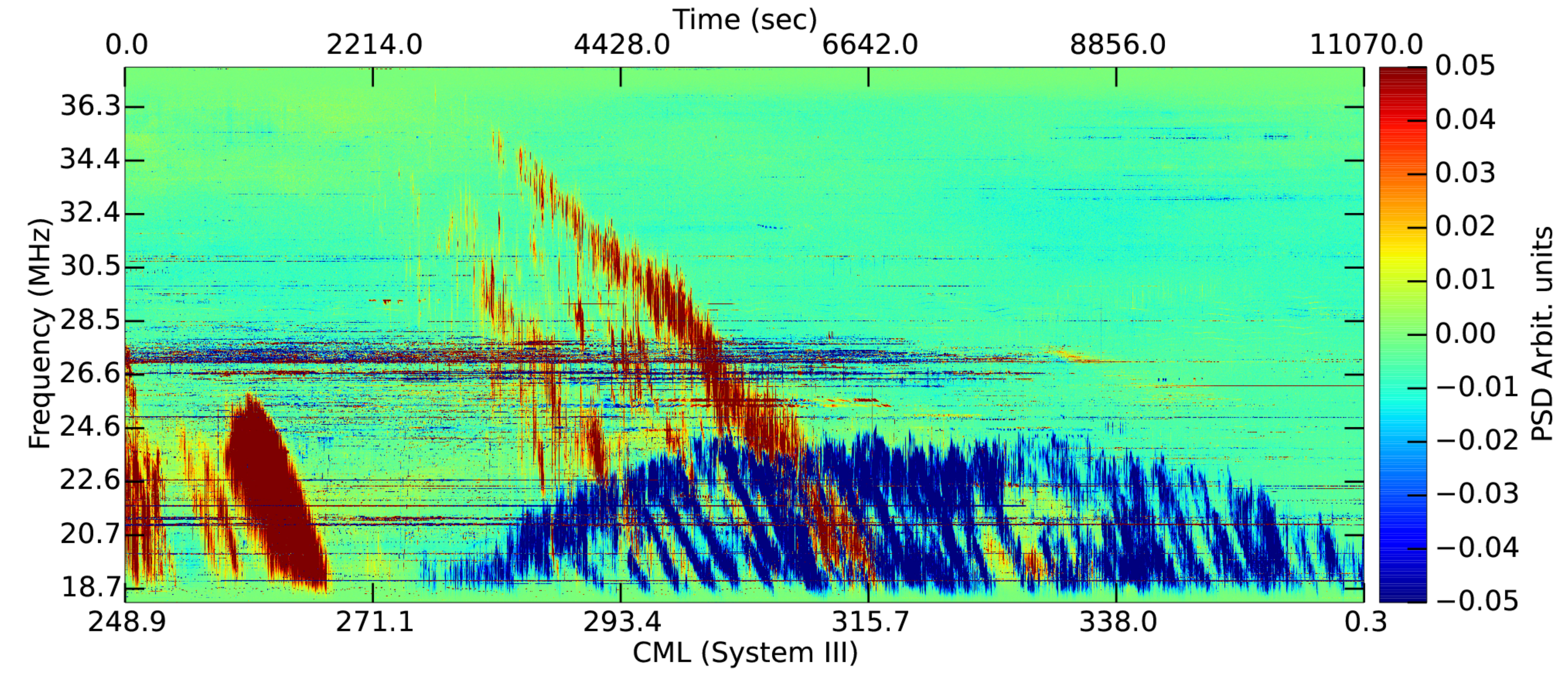}
\caption{Io-A/Io-C burst recorded 06-Mar-2013 for a total duration of
  3.1 hours. Only the upper tuning (centered at 28 MHz) is shown as
  the observations were taken during the daytime and the lower tuning
  was severely impacted by interference. During the early parts of the
  burst the RHC Io-A emission is seen and later the LHC Io-C emission
  is visible. Between \mbox{CML\,{\sc iii}} $\sim$ 280$^\circ$ to
  316$^\circ$ we see RHC emission similar to that in
  Figure~\ref{fig:262}.}
\label{fig:357}
\end{figure*}

The third Io-C event we present, Figure~\ref{fig:357}, was observed on
06-Mar-2013 for a total of 3.1 hours. The early portion of the
observation shows the Io-A emission in RHC ending around
\mbox{CML\,{\sc iii}} = 268$^\circ$ while the LHC emission of the Io-C
event starts around \mbox{CML\,{\sc iii}} = 277$^\circ$ and runs past
the end of our observing window. As with the other two events
discussed above, the Io-C emission shows the vertex late structure
typical of these events. We measure a peak frequency of just above
24.6 MHz for the LHC Io-C emission. Similar to the other two events,
we also see the great arc RHC emission co-incident with the Io-C event
between \mbox{CML\,{\sc iii}} of roughly 275$^\circ$ and 318$^\circ$
before it drops below our upper tuning frequency. Although the lower
tuning is significantly impacted by RFI, we are able to continue
tracing all RHC and LHC emission to frequencies near 12 MHz.

\vfill
\eject

Comparing the three events in Figures~\ref{fig:262} to
\ref{fig:357} we see that the overall picture is one of the Io-A
event ending before \mbox{CML\,{\sc iii}} = 270$^\circ$ followed by
a RHC event which proceeds from higher frequencies (35 MHz) around
\mbox{CML\,{\sc iii}} = 270$^\circ$ to lower frequencies reaching
\mbox{CML\,{\sc iii}} up to 350$^\circ$ before falling below our
lowest frequency observing window. More surprisingly, the LHC Io-C
events appear to start at \mbox{CML\,{\sc iii}} as early as
230$^\circ$ at low frequencies (11 MHz), peaking at roughly 25 MHz
near \mbox{CML\,{\sc iii}} of 300$^\circ$ to 315$^\circ$ before
falling again to frequencies as low as 11 MHz near \mbox{CML\,{\sc
    iii}} of 345$^\circ$. The most recent event reported here (06-Mar-2013) appears
to have the Io-C emission shifted in \mbox{CML\,{\sc iii}} by at least
20$^\circ$ compared to the event observed one year earlier. The
typical \mbox{CML\,{\sc iii}} range quoted for Io-C events is
280$^\circ$ - 60$^\circ$ \citep{carr83}. 

Observations of Io-A/Io-C storms with the NDA are suggestive of an
early start to the Io-C\footnote{Examples can be found in the NDA
  catalog:\\ http://typhon.obspm.fr/maser/SILFE}, but there has not been
any discussion or publications regarding the boundaries of the Io-C
source.  In panel b of Figure 1 in \citet{boudjada91} we see the RHC
and LHC emission from an event on 15-Oct 1987 that shows structure
remarkably similar to the events presented herein in
Figures~\ref{fig:262} to \ref{fig:357}. \citet{boudjada91} note the
occurrence of the LHC Io-C emission in the same \mbox{CML\,{\sc iii}}
range as the RHC but do not trace the early start due to significant
RFI present in the data below 20 MHz.

Polarization studies of Jovian decametric emission using Voyager
planetary radio astronomy (PRA) data by \citet{alexander81} reported a
change in the handedness of circular polarization of continuous DAM
emission below 15 MHz. \citet{ortega84} showed that these anomalies
and sign reversals in the polarization of the PRA below 15 MHz can be
explained by the antenna response when the dipole length is close to
the observing frequency. Using a model for the polarization response
for the PRA, \citet{ortega91} were able to analyze the polarization
properties of the hectometric emission and separate the contributions
of two main sources of emission. They found these two sources had
opposite handedness and independently changing intensities. Due to the
complexities of the higher frequency PRA band, they were not able to
map the polarization details of the DAM. At these higher frequencies
from Voyager only total intensity analyses of Jupiter's DAM were made
and thus it would have been very difficult to detect the early start
of the Io-C emission we observe with the LWA1.

\begin{figure*}[th!]
\centering
\noindent\includegraphics[width=40pc]{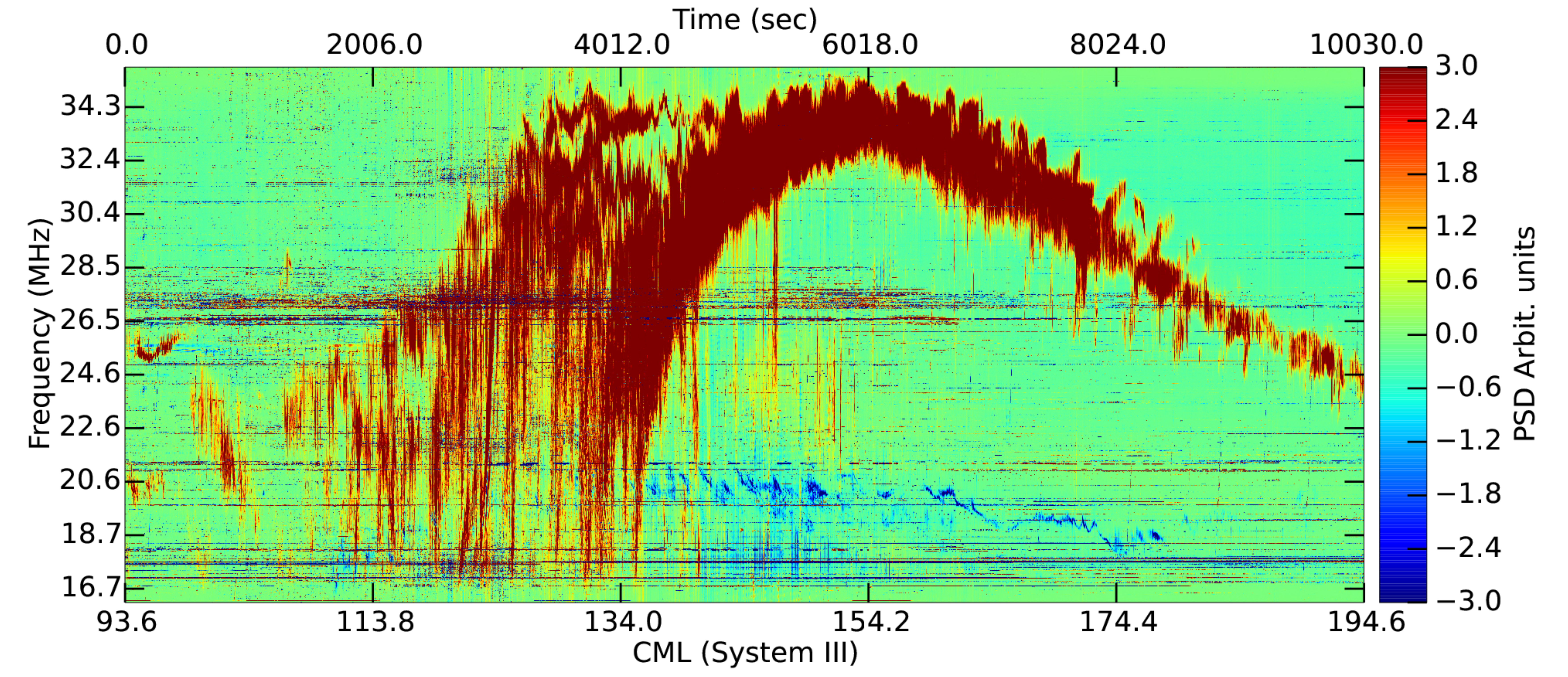}
\caption{Upper tuning (26 MHz) for the Io-B burst recorded on
  11-Mar-2012. The RHC emission from the Io-B event peaks at
  frequencies near the upper end of our observing window ($\sim$ 35
  MHz) and is seen to cover the full observing time.
  We also see faint, narrow-band LHC emission peaking near 21 MHz which
  we interpret as Io-D emission.}
\label{fig:997_full}
\end{figure*}

\subsection{Io-B}
\label{sect:Io-B}

Io-B events are generally considered as originating in the northern
hemisphere and are thought to represent the leading edge of the
northern emission cone. In Figure~\ref{fig:997_full} we show the upper
tuning (centered at 26 MHz) of a classic Io-B observation from
11-Mar-2012.  Early in the event we see emission drifting upward in
frequency and peaking about 35 MHz followed by a more narrow-band tail
that drifts downward in frequency. The Io-B source shows vertex early
arc structure that have been well studied and modeled \citep{hess08,
  queinnec98, wilkinson89, leblanc81, gs81}.  As is common in Io-B
events, S-bursts are seen in the early part of the event. We also see
a faint, and more narrow-band LHC emission ranging from
\mbox{CML\,{\sc iii}} values of about 130$^\circ$ to 180$^\circ$.
This emission peaks about 21 MHz and is most likely southern
hemisphere Io-D emission.  We discuss S-bursts in
Section~\ref{sect:Sbursts} and the Io-D emission in
Section~\ref{sect:Io-D}.

\begin{figure*}[bh!]
\centering
\noindent\includegraphics[width=40pc]{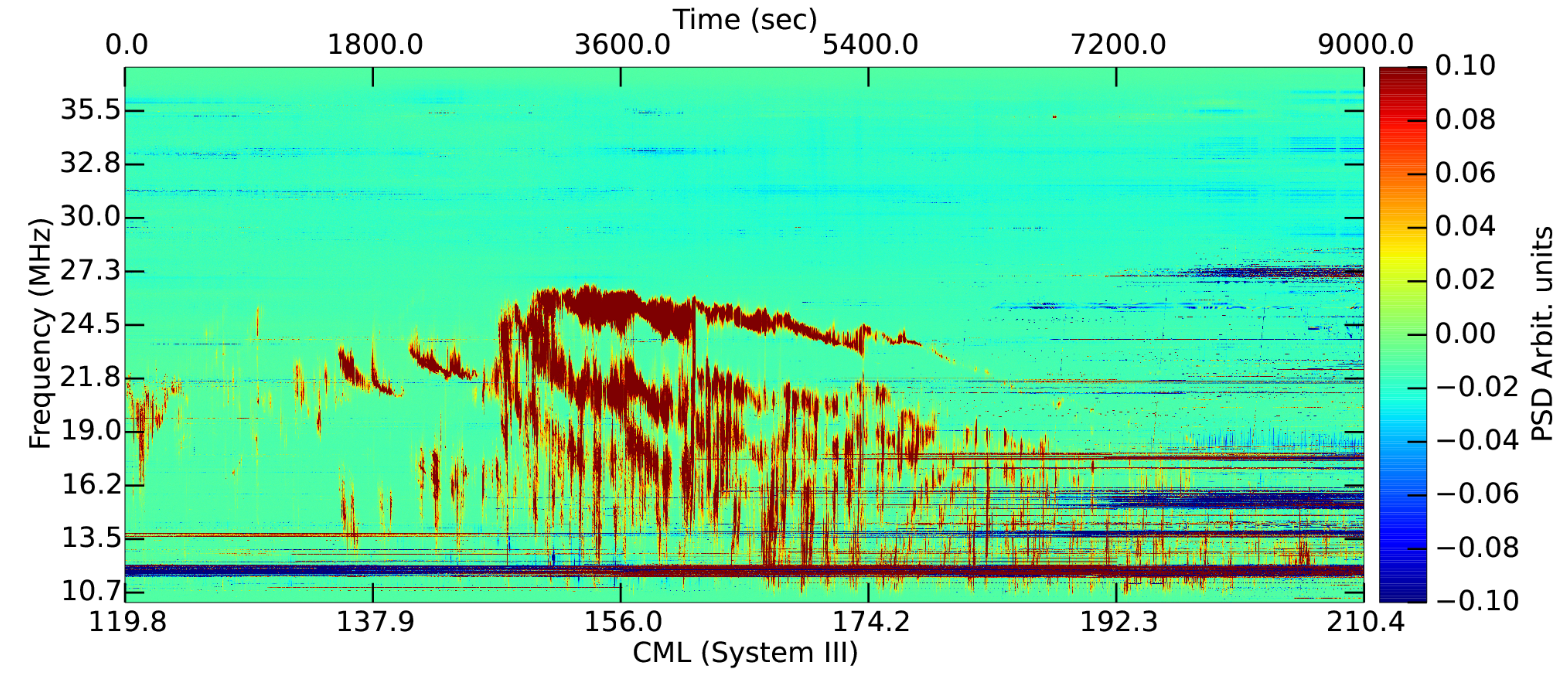}
\caption{Narrow band Io-B burst recorded 24-Sept-2012. The burst is
  comprised of narrow bands of S-bursts and N-events and is host to a
  variety of features including the Tilted-V events described in
  \citep{Riihimaa90}. }
\label{fig:194_full}
\end{figure*}

A very different Io-B event showing a series of narrow-banded emission
regions is shown in Figure~\ref{fig:194_full}. This observation covers
a total of 2.5 hours. This entire event consists of narrow-band
emission bands. Analysis of the high temporal resolution spectrograms
of this burst show a number of fine structure details such as S bursts
and Tilted V events. These latter events are seen as regions within
narrow-band events that are bounded by two S bursts \citep{Riihimaa90}
and are discussed further in Section~\ref{sect:Sbursts}.

\begin{figure*}[tb!]
\centering
\noindent\includegraphics[width=40pc]{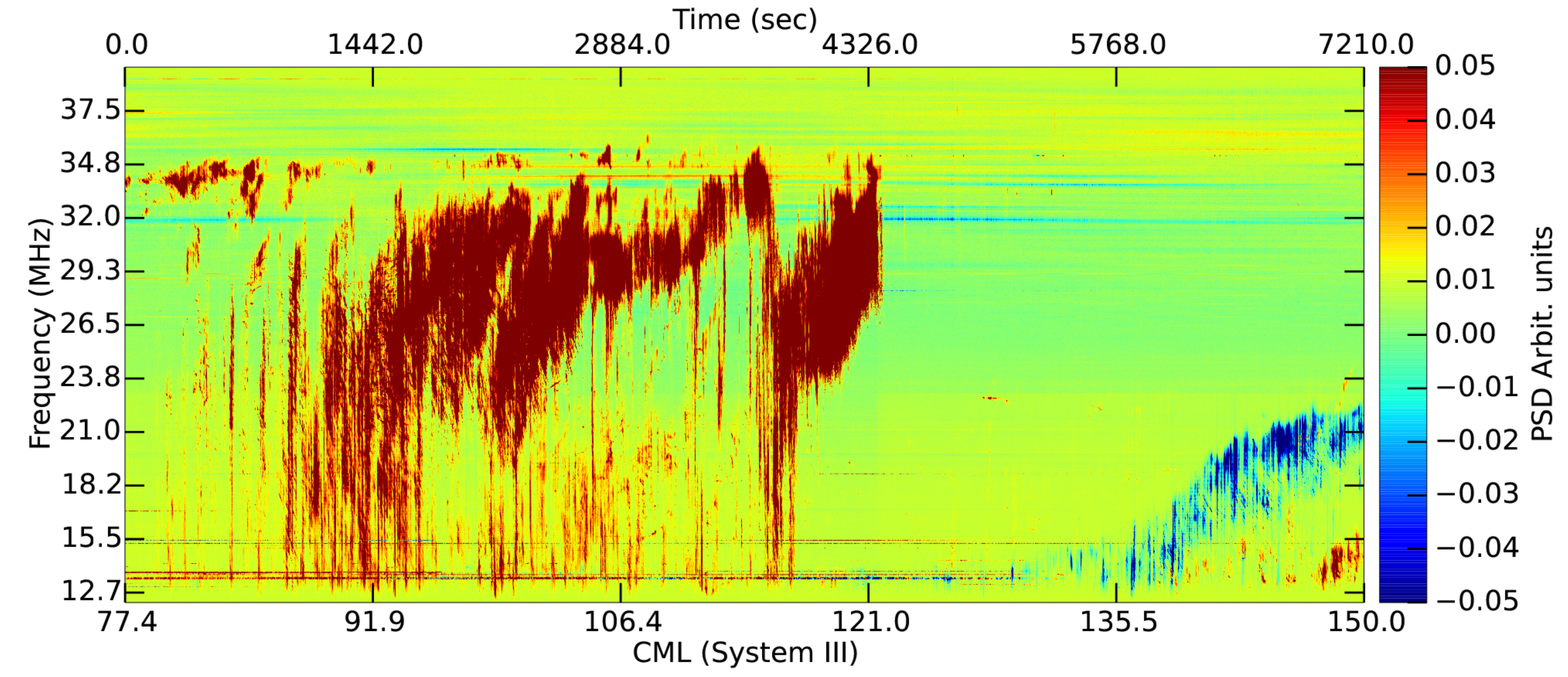}
\caption{Io-B event from 27-Dec-2012 which shows
  vertical arc structures and a maximum frequency of 35 MHz. A sharp
  cutoff of the emission occurs at (\mbox{CML\,{\sc iii}} 121$^\circ$,
  Io Phase 100$^\circ$), just as the source exits the Io-B region (see
  Figure~\ref{fig:prob}). At later times, Io-D emission is clearly evident beginning about
  (\mbox{CML\,{\sc iii}} 130$^\circ$) and reaches a peak frequency of 22.5 MHz.}
\label{fig:288_full}
\end{figure*}

The final Io-B/Io-D event we present is shown in
Figure~\ref{fig:288_full}. This event is peculiar because the Io-B
emission cuts off sharply at 07:43 UT (\mbox{CML\,{\sc iii}}
121$^\circ$, Io Phase 100$^\circ$), presumably just as the source cone
rotates past the Earth. Because the LWA1 off-beam data (not shown
here) show good continuity over this interval, we believe that this
sharp cutoff is real and displays the narrow beaming nature of the DAM
emission cones.  Figure~\ref{fig:prob} shows that this event was a
high pass through the Io-B probability region on the Io
phase-\mbox{CML\,{\sc iii}} plane. Soon after the Io-B cone moves out
of our line of sight an Io-D event begins. This event is seen as the
LHC emission. Unfortunately this observation ended at 09:40 UT
(\mbox{CML\,{\sc iii}} 150$^\circ$, Io Phase 107$^\circ$) just as Io-D
appeared to be reaching its peak frequency. The maximum frequency
observed before the end of the observations was 22.5 MHz. This Io-D
event is also particularly interesting due to the detection of
S-bursts within it (see Section~\ref{sect:D_FS}).

\subsection{Io-D}
\label{sect:Io-D}

The leading edge of the southern emission cone is thought to be
associated with Io-D events. These sources are more elusive due to the
lower emission intensity, lower emission frequency and the overlap of
the source emission region with the same general \mbox{CML\,{\sc iii}}
and Io Phase ($\Phi_{Io}$) range as the stronger Io-B radiation. The
\mbox{CML\,{\sc iii}} and Io phase ranges for the Io-D source are
\mbox{CML\,{\sc iii}} = 0$^\circ$ to 200$^\circ$ and $\Phi_{Io}$ =
80$^\circ$ to 130$^\circ$ \citep{carr83}. \citet{queinnec98} found
that the Io-D reaches down to frequencies of 3 MHz and that it appears
as an isolated arc. Their model argues for the Io-D source in the
southern hemisphere along a single Io flux tube. We note that we did
not specifically target Io-D events during our initial observing
program but several events were detected within the targeted Io-B
event observations. Based on these detections, we have initiated a
follow-up observing program which specifically targets the Io-D
sources for detailed structure analysis and those results will be
presented in a future paper.

In Figure~\ref{fig:997_full} we see evidence of weak, narrow-band LHC
polarization extending over \mbox{CML\,{\sc iii}} = 134$^\circ$ to
185$^\circ$ and $\Phi_{Io}$ = 93$^\circ$ to 104$^\circ$. The emission
is slowly drifting in time with the peak frequency moving from a
maximum near 21 MHz to around 18 MHz. The emission characteristics of
this Io-D event are consistent with results from both NDA and
Voyager data \citep[see][]{leblanc81, hess08}. Structures identified
in higher time-resolution images of this LHC emission are discussed in
Section~\ref{sect:D_FS}.

The second Io-D event we present can be seen beginning near the end of
the 27-Dec-2012 event shown in Figure~\ref{fig:288_full}. Shortly
after the Io-B emission cone passes out of our line-of-sight, the LHC
Io-D emission begins. The emission is initially at very low
frequencies ($\sim$ 13 MHz) near \mbox{CML\,{\sc iii}}
121$^\circ$. The emission frequency increases until the end of our
observations at which point the peak frequency is near 22 MHz. Due to
the remarkably clean radio frequency interference spectrum at these
low frequencies we are able to see evidence of both Riihimaa
modulation lanes as well as nearly horizontal, closely spaced Faraday
lanes that are clearly seen below 21 MHz across both the Io-B and Io-D
emission regions. We show these Faraday fringes in
Figure~\ref{fig:288_FL} and discuss this further in
Section~\ref{sect:Faraday}. High time-resolution study of this Io-D
burst shows clear S-burst structure within the event. In
Section~\ref{sect:D_FS} we discuss the rare Io-D fine structure and
characterize the drift rate of these S-bursts.

\section{Modulation Lanes and Fine Structure}
\label{sect:fine_structure}

There are many examples of fine structures clearly seen in the LWA1
data.  Most of these structures have been seen by other ground-based
and spacecraft observers \citep{litvinenko09, zarka98, carr83} in one
or more Io-controlled events. Examples include frequency envelopes for
different sources, nested arc features, narrow band events, modulation
lanes, Faraday lanes, and simple and complex S-bursts. We will
highlight many of the excellent LWA1 observations and address these
features in the sections below.

\subsection{Modulation Lanes}
\label{sect:modulation}

Jovian emission undergoes a variety of propagation effects as it
passes from the emission cone near the planet through the magnetoionic
plasma in the Io torus, the interplanetary medium and the Earth's
ionosphere. \citet{Riihimaa68} identified a set of amplitude
modulation lanes that cut across the L-bursts and S-bursts and which
could have either positive or negative slopes. These modulation lanes
are characterized by a strong periodicity in time, with periods
ranging from about 1 to 5 sec and an average of about 2 sec. 

\begin{figure*}[tbh!]
\centering
\noindent\includegraphics[width=40pc]{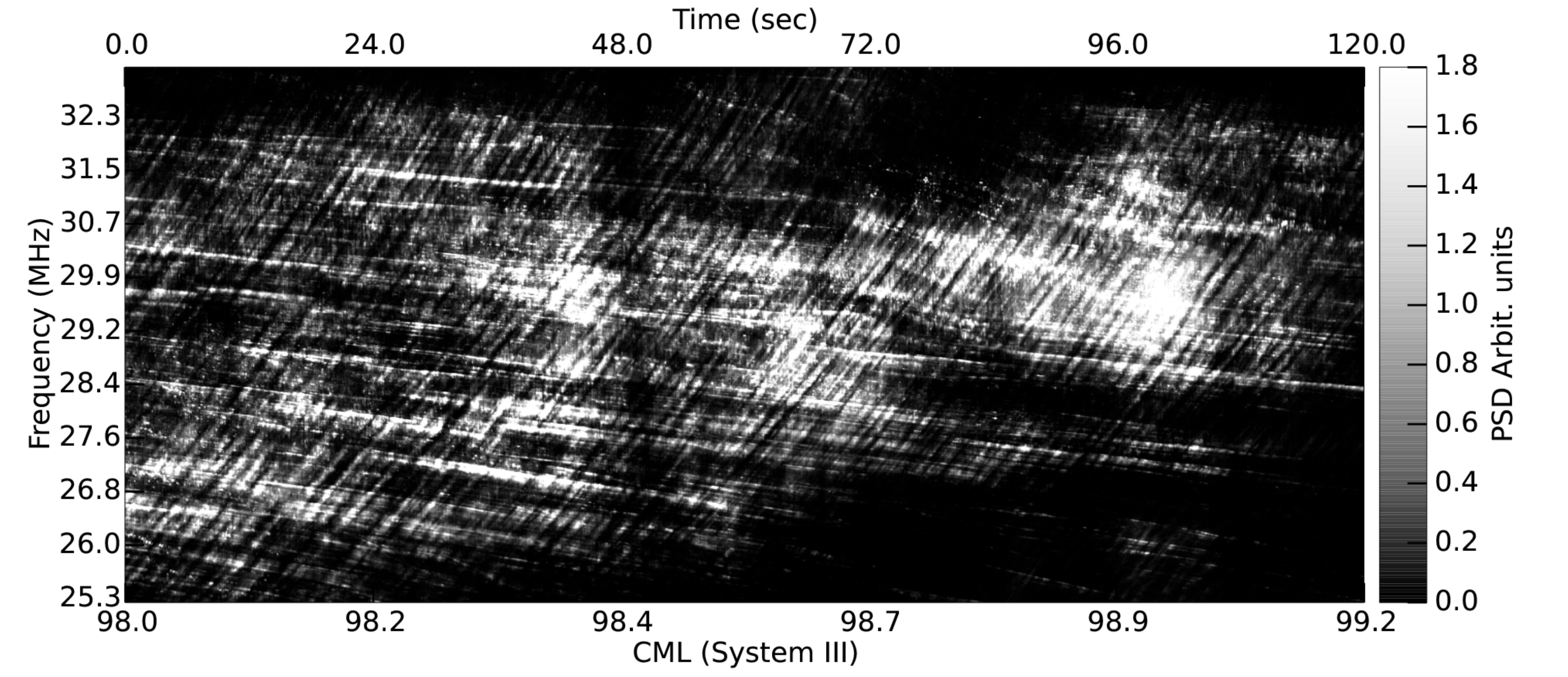}
\caption{Positive slope modulation lanes seen in a portion of the Io-B
  burst of 27-Dec-2012 which is shown in
  Figure~\ref{fig:288_full}. The image is made with a temporal
  resolution of 10 ms and spectral resolution of 10 kHz. }
\label{fig:288_mod_lanes}
\end{figure*}

In Figure~\ref{fig:288_mod_lanes}, we show a small 120 second portion
of the Io-B event shown in Figure~\ref{fig:288_full} where positive
slope modulation lanes are clearly visible. This image has a temporal
resolution of 10 ms and spectral resolution of 10 kHz. The modulation
lanes show regular spacing crossing the Io-B emission. Slope
measurements of the lines in the frequency range of 21-23 MHz in early
parts of this 27-Dec-2012 event show slopes consistent with
measurements from \citet{Riihimaa68}.
\begin{figure*}[th!]
\centering
\noindent\includegraphics[width=40pc]{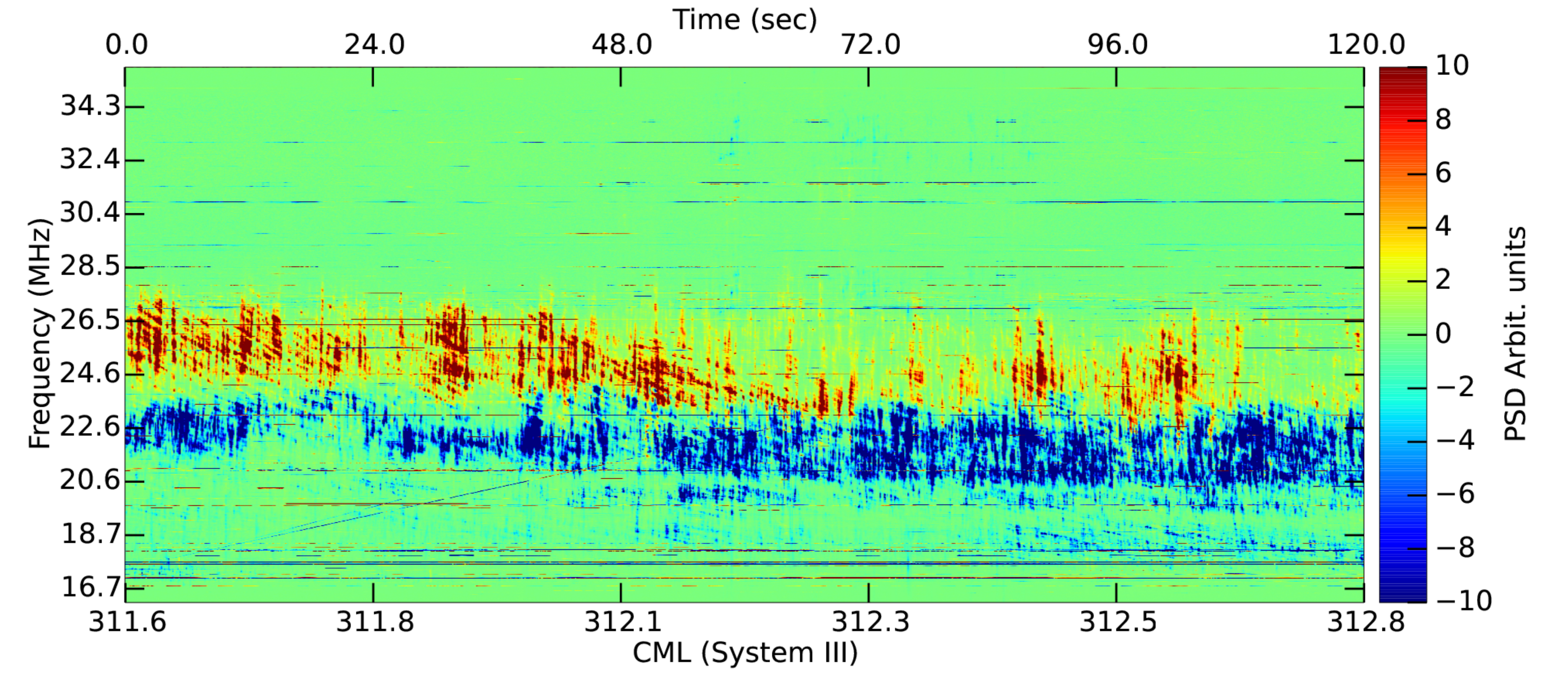}
\caption{Zoom in of the Io-A/Io-C burst recorded on 10-Mar-2012
  showing 120 seconds of data starting at 23:16:41 UT. Modulations
  lanes are visible as negative drifting features (decreasing
  frequency with time) running through the LHC and RHC emission. The
  two LHC features running from low to higher frequencies seen from
  the beginning of the observation to the middle are ionospheric
  sounders. Color bar as in Figure~\ref{fig:262}.}
\label{fig:mod_lanes}
\end{figure*}

High temporal resolution spectrograms of the 10-Mar-2012 event show
modulation lanes visible through the RHC and LHC emission of this
event (Figure~\ref{fig:mod_lanes}). Analysis of these lanes shows that
they are continuous across the two different polarization hands
\citep{imai13}.  We defer a detailed analysis of this modulation lane
structure to a future paper but do note that the detection of these
modulation lanes may have very important implications for our
understanding of the emission mechanism at Jupiter. If both the RHC
and LHC emission regions are shown to be crossed by continuous
modulation lanes then, according to the widely accepted \citet{imai92}
model, this would mean that the two polarization hands must be coming
from sources in the {\it same} hemisphere. This would then suggest
that the emission detected would be both the L-O and R-X modes of the
CMI. One concern is that theoretical studies of the CMI emission
mechanism of Earth's auroral kilometric radiation (AKR), which is
analogous to Jovian DAM emission, predict that the R-X mode growth
rates up to two orders of magnitude larger than the L-O mode
\citep{melrose84}. Observational studies of AKR support these
predictions \citep{hanasz03}.  Our data show roughly similar powers
for both hands of polarization which may imply that other processes
are necessary to explain the L-O emission, like refractions,
reflections, or mode conversions.  Alternately, if the emissions are
coming from opposite hemispheres, the propagation paths of each source
emission through an interference screen that is comprised of
field-aligned columns of enhanced plasma density from the Io torus
would need to be remarkably similar to form the continuous modulation
lanes across both polarizations \citep[see][]{imai97}.

\begin{figure}[h!]
\noindent\includegraphics[width=3.5in]{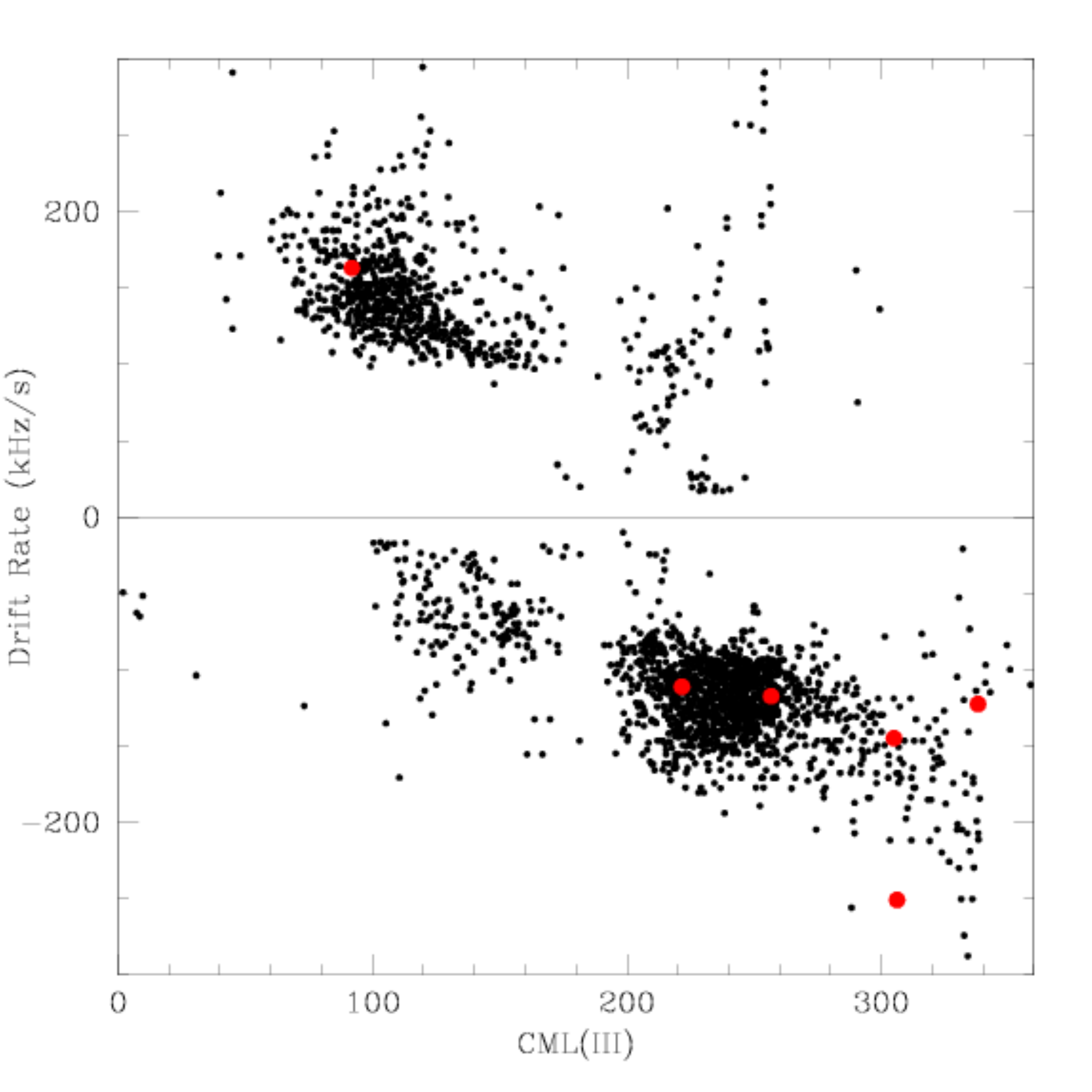}
\caption{Modulation lane slope vs \mbox{CML\,{\sc iii}} plotted for
  data from \citet{Riihimaa68} in black with measurements from the
  LWA1 bursts shown as red points. Each LWA1 slope measurement is the
  mean of 6 independent slope measurements taken in nearly adjacent
  lanes around the \mbox{CML\,{\sc iii}} value. Each point is measured
  for lanes in a single burst but multiple bursts were used for the
  range of measurements shown. }
\label{fig:slope_cml}
\end{figure}

We have measured the slope of the modulations lanes for the above LWA1
events. We show in Figure~\ref{fig:slope_cml} the measured slopes of
several of the LWA1-detected modulation lanes included on the
modulation lane plot of \citet{Riihimaa68}. Each LWA1 slope is the
mean of independent slope measurements for 5 consecutive modulation
lanes around the plotted \mbox{CML\,{\sc iii}}. The figure shows the
modulation lane slopes we measure are consistent with earlier papers
although we note that we do detect modulation lanes in Io-C regions of the
plot that are currently sparsely populated. We defer a detailed
modulation lane slope analysis with many more measurements to a future
paper.  
 
\subsection{Faraday Fringes}
\label{sect:Faraday}

Highly elliptically polarized emission passing through dense plasma
media, such as Jupiter's Io plasma torus plus magnetosphere and the
Earth's ionosphere, will undergo Faraday rotation due to the
bi-refringent nature of the magnetized and ionized plasma. 

Each magneto-ionic medium between the source and observer will
introduce a rotation measure (RM) in the polarized Jovian emission
where \citet{tools} show the Faraday rotation measure along some path
$\ell$ is defined as:
\begin{equation}
RM = \frac{e^3}{2\pi m_e^2 c^4} \int_0^\ell n_e(s)B_\|(s)\ ds \ \  {\rm radians/m^2}
\end{equation}
where $e$ is the electron charge, $m_e$ is the mass of the electron, $c$ is the speed of light, 
$n_e$ is the electron density at each point $s$ along the path, and $B_\|$ is line of sight 
component of the magnetic field at point $s$. 

\begin{figure*}[tbh!]
\centering
\noindent\includegraphics[width=40pc]{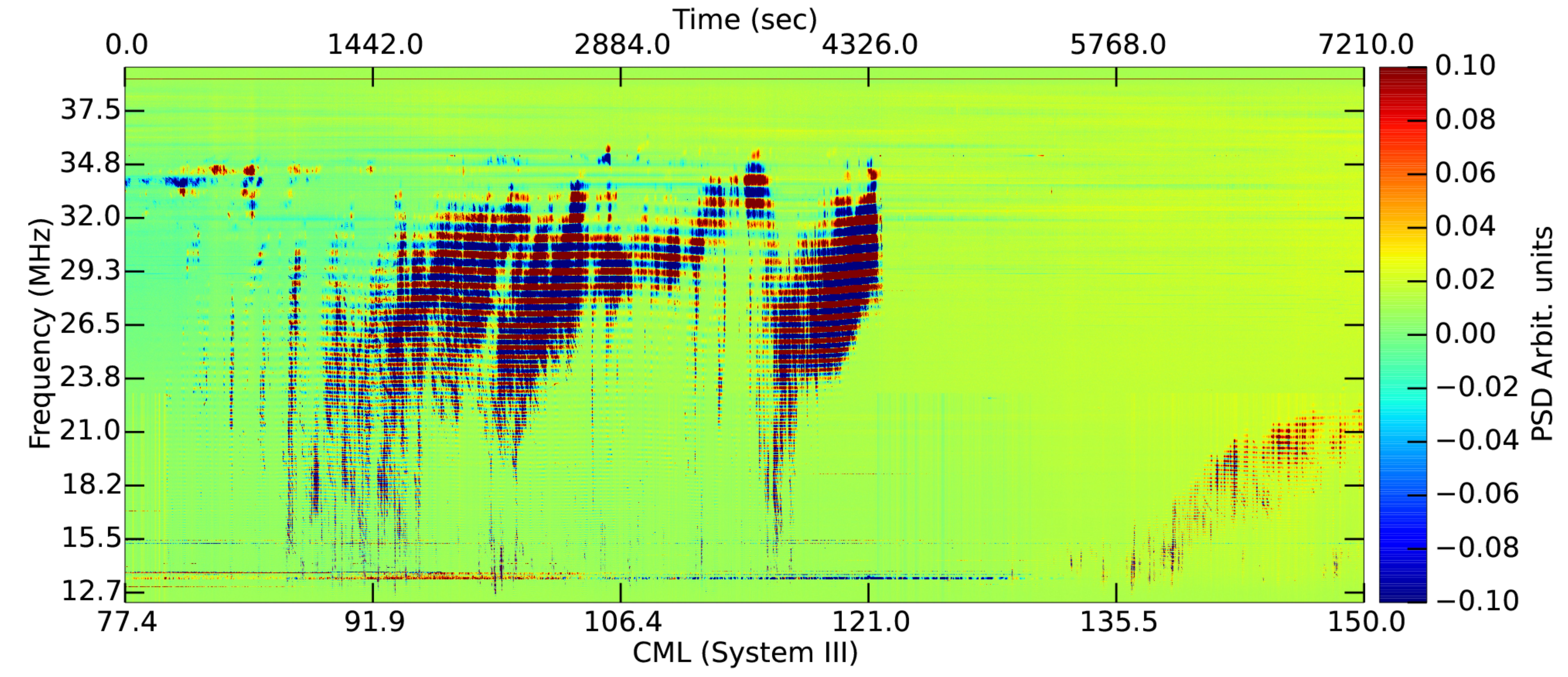}
\caption{Plot of 27-Dec-2012 burst showing Faraday fringes in a linear
  polarization difference spectrogram. The nearly horizontal Faraday
fringe bands are seen across the entire Io-B and Io-D emission
regions. The fringe separation decreases as the frequency decreases
due to the $\lambda^2$ nature of Faraday rotation.}
\label{fig:288_FL}
\end{figure*}

There are at least four distinct contributions to the observed Faraday
rotation of Jovian decametric emission. These are (1) the Jovian
magnetosphere, (2) the Io plasma torus, (3) the interplanetary medium,
and (4) the Earth's ionosphere. Each of these components will make a
time-variable contribution to the observed Faraday rotation.  Taking
into account the propagation effects from these media, \citet{dulk92}
used Io-B events observed from the NDA to
determine that at least some of the Faraday rotation comes from the Io
plasma torus. The fraction of Faraday rotation from each of the
components above is unclear from work in the literature and is
discussed in some detail in \citet{litvinenko09}. The majority of the
RM signal likely comes from the Earth's ionosphere which has typical
electron density of $n_e = 5 \times 10^5$ cm$^{-3}$, field strength of
$B_\| = 0.3$ G, and a path length of 500 km. This results in a
contribution of roughly RM$_\oplus \sim 2$ rad/m$^2$.

Linearly polarized feeds, such as used at LWA1, will detect the
emission with a time-variable visibility in each of the two orthogonal
polarizations. We have investigated several of the bursts using the
orthogonal linear polarizations from the LWA1 and see clear evidence
of Faraday fringes. We show in Figure~\ref{fig:288_FL} a linear
polarization difference spectrogram of a portion of the 27-Dec-2012
Io-B/Io-D event with clear nearly horizontal Faraday fringes. Similar
Faraday fringes have been seen for the Io-B event on 11-Mar-2012. We
have used the locations of the nulls in the latter event to determine
the rotation measure of RM=2.2 rad m$^{-2}$ which is similar to
the expected contribution from the Earth's ionosphere. 

\subsection{Fine Structure}
\label{sect:B_sbursts}

In this section we discuss some of the fine structure seen in the LWA1
observations of the Jovian bursts. Short bursts of emission, known as
S-bursts, are described in detail in \citep{Zarka96, carr83}. These
sources consist of a series of very short pulses of emission that
range in length from 1 ms to several hundred ms. The instantaneous
emission is very narrow in frequency and typical drift rates are
between -5 and -30 MHz/s \citep{Zarka96}. Drift rates have been
measured for S-burst seen in Io-B and Io-C events, and more rarely in
Io-A events \citep{hess09,flagg}. Assuming that the emission occurs
  near the local electron cyclotron frequency, the negative slope is
  interpreted as resulting from motion of the packet of radiating
  electrons away from the Jovian surface toward lower magnetic field
  strengths.  \citet{carr99} show the fine structure within an S-burst
  and the coherence of the oscillations inside the microstructure
  while wavelet analysis of these bursts shows modulations on scales
  of 6-15 $\mu$s \citep{litvinenko04}.

N-events are described in \citet{riihimaa85}. These events are
relatively rare and consist of narrow frequency bands with trains of
emission that can drift up or down with time. The N-events are often
either composed of short S-bursts of multiple 'split' bands of
emission. Typical bandwidths are 200 - 600 kHz with durations of a few
seconds to tens of minutes. These events are also seen to drift slowly
in frequency with time in some cases. \citet{Riihimaa68} studied the
occurrence of the N-events and found they are often seen over a wide
range of \mbox{CML\,{\sc iii}}.

\subsubsection{Io-B Fine Structure}
\label{sect:Sbursts}

\begin{figure*}[tbh!]
\centering
\noindent\includegraphics[width=40pc]{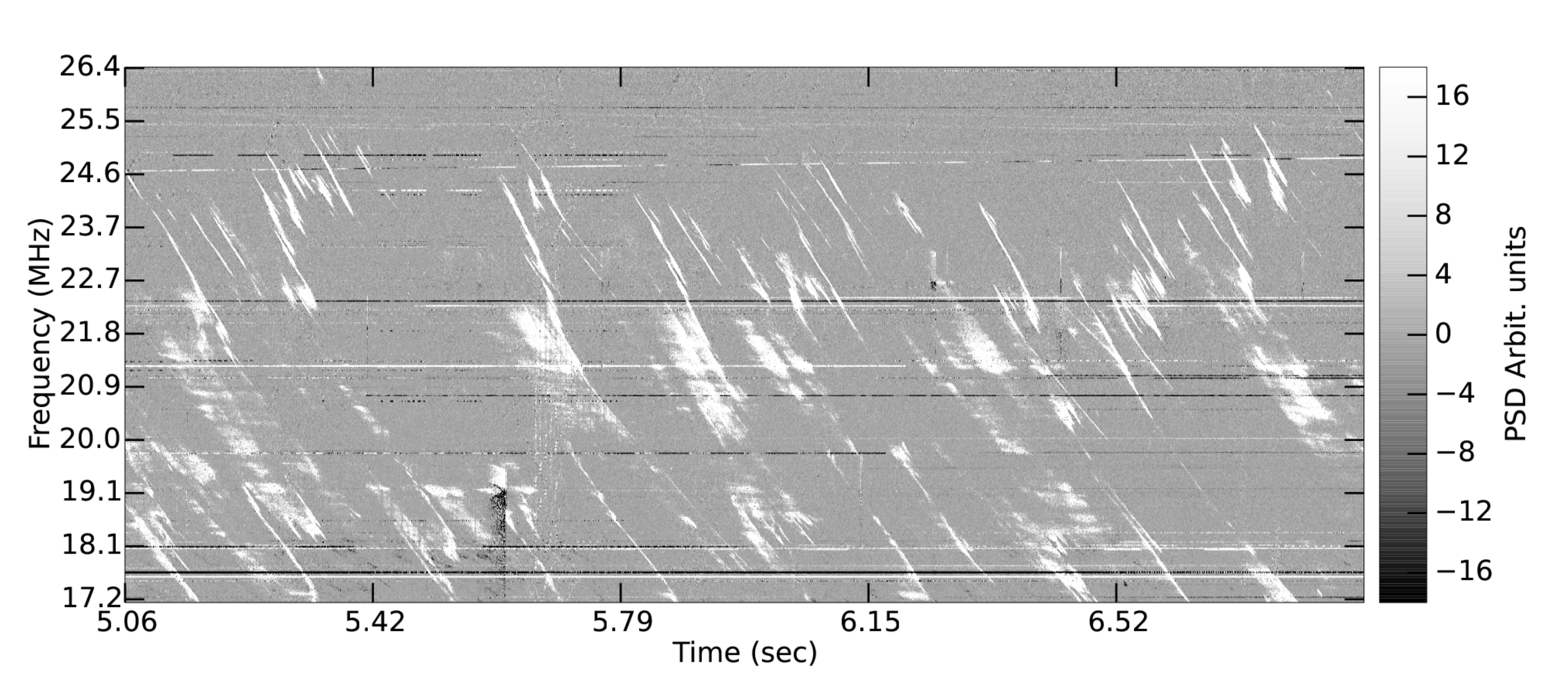}
\caption{Multiple S-bursts plotted at 0.25 ms and 10 kHz resolution
  over 1.825 seconds of the 11-March-2012 Io-B event. The S- and
  N-events show apparent destructive interaction of the two bursts.}
\label{fig:IoB_SN_wide}
\end{figure*}

We have investigated the fine structure of the 11-March-2012 Io-B
event (Figure~\ref{fig:997_full}) at high temporal
resolution. Figure~\ref{fig:IoB_SN_wide} shows a short $\sim$ 1.8 s
long portion of the burst with 0.25 ms temporal resolution and 10 kHz
spectral resolution for a 9.2 MHz wide frequency window centered
around 21.8 MHz. We see that the portion of
the spectrum between 21 and 25 MHz is largely dominated by S-bursts
while below 21 MHz there is a mix of S-bursts and N-events. We have
measured the slopes of the S-bursts in several regions of the Io-B
event and find values of around -15 MHz/s, consistent with typical
values for Io-B events reported in the literature \citep[see Figure 3
  in][]{Zarka96}.

\begin{figure*}[bh!]
\centering
\noindent\includegraphics[width=40pc]{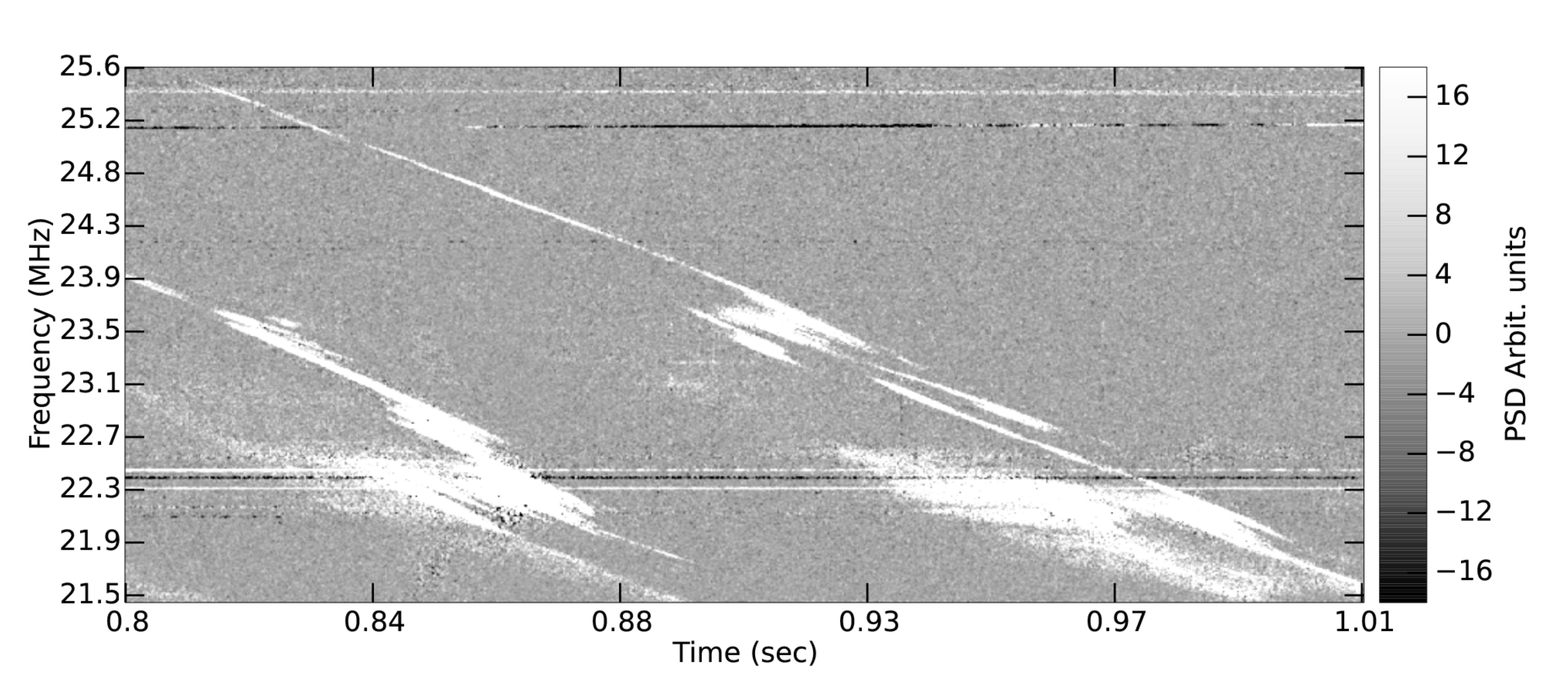}
\caption{Zoom in on a single S-burst at 0.25 ms temporal resolution
  and 10 kHz spectral resolution over a total of 0.21 s during the 11-March-2012 Io-B event. }
\label{fig:IoB_SN_zoom}
\end{figure*}

More complex S-burst interactions with N-events are seen below 23 MHz
in Figure~\ref{fig:IoB_SN_wide} where there appears to be multiple
instances of N-events terminating at the time when an S-burst passes
through the same frequency as the N-event.  In
Figure~\ref{fig:IoB_SN_zoom} we isolate details of one such
interaction over a total time of 0.21 s. This figure shows the full
details of one S-burst that starts near 25.6 MHz and drifts at a rate
of -17.9 MHz/s to roughly 23.7 MHz where it encounters and apparently
quenches an N-event which began $\sim$ 15 ms before the S-burst
encounter. Multiple S-bursts appear to be generated after the
termination of the N-event starting at frequencies of 23.2 MHz and
with slopes of -16.3 MHz/s. Analysis of high temporal resolution data on
this event appears to show that there are several regions where the
S-bursts and short N-events are apparently interacting. This would
suggest that the source regions for each type of event are cospatial.

\begin{figure*}[t!]
\centering
\noindent\includegraphics[width=40pc]{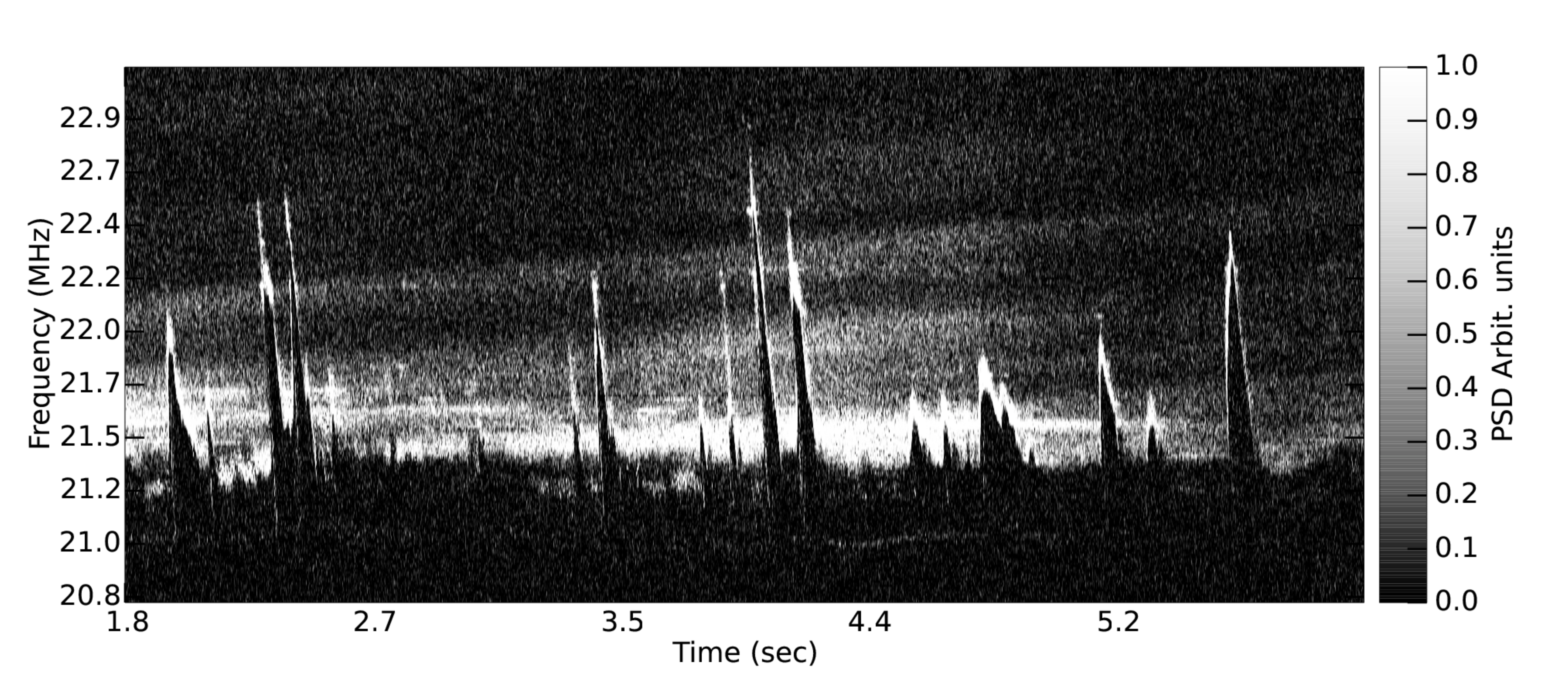}
\caption{This spectrogram, shown in black-and-white for clarity, shows
  one of the narrow bands of Io-B emission during the 24-Sept-2012
  event seen in Figure~\ref{fig:194_full}.  More than a dozen inverted “tilted-V”
  shaped narrow band emissions are seen in this 4.3 second
  spectrogram. The emissions have a baseline frequency of about 21.3
  MHz and are seen to fluctuate rapidly to frequencies as high as 22.7
  MHz and back over 0.1 seconds. Also seen are the “shadow event”
  regions of near zero emission beneath the tilted-V shapes.}
\label{fig:194_tiltv}
\end{figure*}

Previous observations have been made that show interactions between
N-events and S-bursts, referred to as S-N burst events \citep{oya02,
  riihimaa85, riihimaa81}. The S-N burst events typically occur
between 20-24 MHz and show an inverted “tilted-V” shape in the
spectrum that connects the S-burst with the N
event. Figure~\ref{fig:194_tiltv} shows a 4.3-second portion of one of
the narrow bands of emission during the 24-Sept-2012 narrow-band Io-B
event (see Figure~\ref{fig:194_full}). The spectrogram in
Figure~\ref{fig:194_tiltv} has a bandwidth of 2.3 MHz centered on 22
MHz and clearly shows many inverted-V shapes as narrow band emissions
changing rapidly in frequency and in time. Also seen are the “shadow
event” regions beneath the tilted-V shapes. The origin of these events
and interactions are not clearly understood, but are thought to be
bunched electrons accelerated by electric fields upward along the Io
flux tube \citep[cf.][and references therein]{oya02}. Thus these S-N
burst events are believed to be generated in the same region in
Jupiter's magnetosphere; our observations of S- and N-events shown in
Figures~\ref{fig:IoB_SN_wide} and \ref{fig:IoB_SN_zoom} are consistent
with this interpretation. More high resolution observations of the
Io-B source using the LWA instrument may provide some additional clues
as to the nature and origin of these peculiar events.

\subsubsection{Rare S-Bursts in an Io-D event}
\label{sect:D_FS}

Very little information is available in the literature regarding the
Io-D events. In general, the Io-D emission seems to be characterized
by relatively narrow emission seen over a wide range of
\mbox{CML\,{\sc iii}} and confined to frequencies below $\sim$ 18
MHz \citep{carr83}.

\begin{figure*}[b!]
\centering
\noindent\includegraphics[width=40pc]{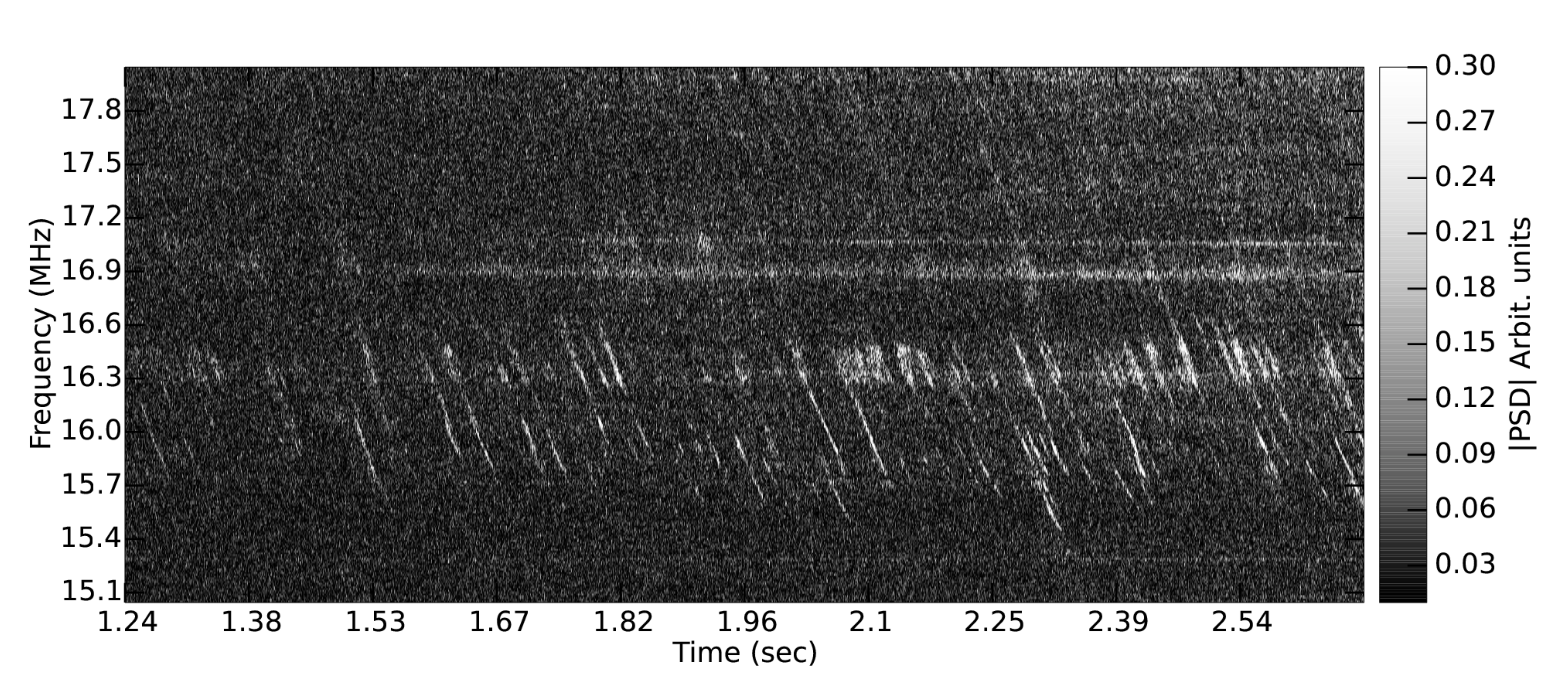}
\caption{S-bursts for the Io-D event on 27-Dec-2012 (see
  Figure~\ref{fig:288_full}) plotted at 0.25 ms temporal resolution
  and 10 kHz spectral resolution. The bursts are LHC but are shown in
  the figure as an absolute value in greyscale to enhance the
  visibility of the bursts. Times are shown relative to 6236 seconds
  into the burst with a total time of 1.44 seconds displayed.}
\label{fig:IoD_sbursts}
\end{figure*}

We have undertaken high temporal resolution studies of the Io-D
emission from the 27-Dec-2012 event and we see clear S-bursts during
this LHC emission event. A short portion of the spectrum ($\sim$ 1.44
s) is shown in Figure~\ref{fig:IoD_sbursts} where we show the region
of the spectrum from $\sim$ 15 MHz to 18 MHz. We measure drift rates
of the Io-D S-bursts of -11 to -15 MHz/s at frequencies of 15 to 16
MHz which are consistent with drift rates from other Io sources. There
are few discussions in the literature regarding S-burst detections
during known Io-D events, and only one definitive event could be found
in the literature \citep[see][]{litvinenko10}. Because both Io-B and
Io-D have significant overlap in CML, it is difficult to distinguish
the Io-related source of the S-bursts. There is some possible evidence
showing Io-D related S-bursts in the Voyager PRA data (see
\citet{alexander84}) and Cassini radio and plasma wave science (RPWS)
data (see \citet{lecacheux01}).

\begin{figure*}[t!]
\centering
\noindent\includegraphics[width=40pc]{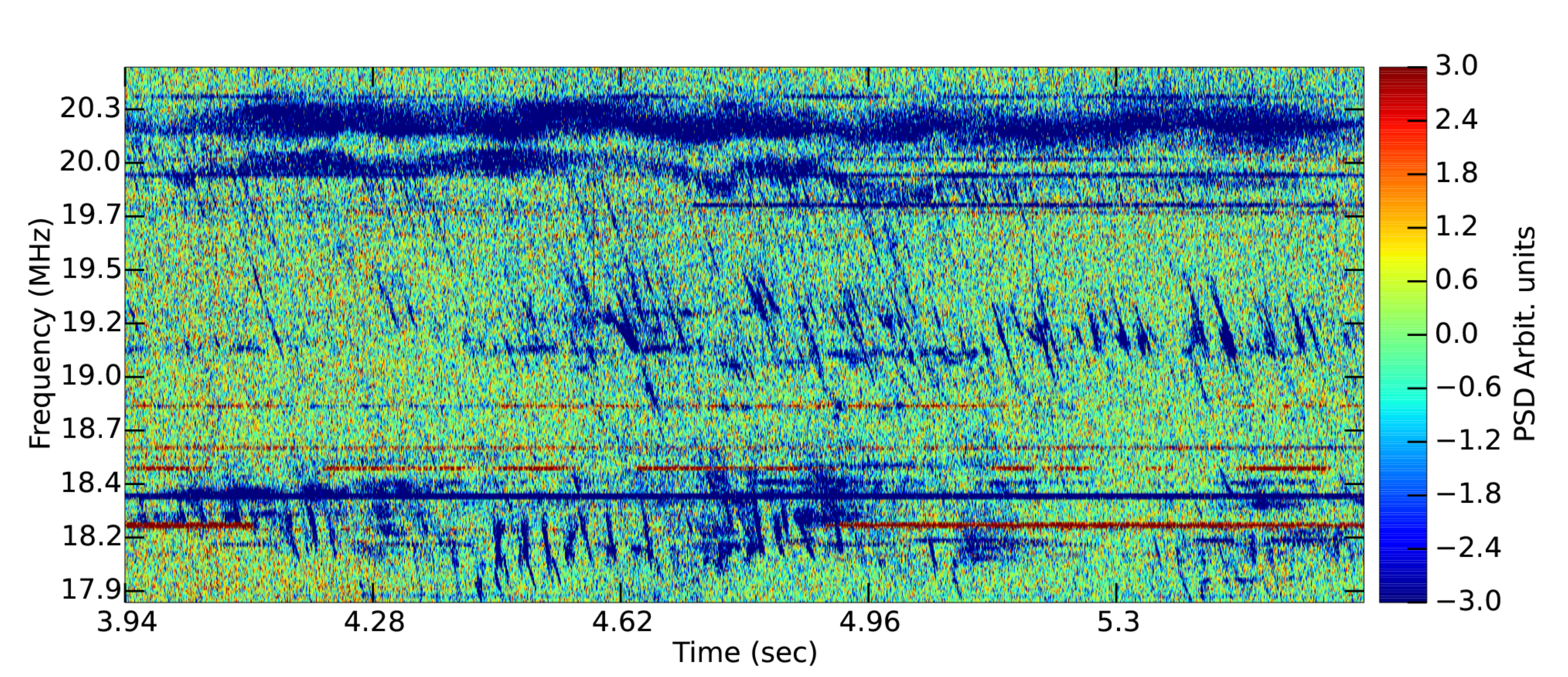}
\caption{S-bursts for the Io-D event on 11-March-2012 (see
  Figure~\ref{fig:997_full}) plotted at 0.25 ms temporal resolution
  and 10 kHz spectral resolution. Narrow-band emission as well as S-bursts
are seen within the LHC Io-D emission.}
\label{fig:IoD_997}
\end{figure*}

In Figure~\ref{fig:IoD_997} we show the details of the LHC Io-D
emission from the 11-March-2012 event with 0.25 ms temporal resolution
and 10 kHz spectral resolution. The figure shows a window of 1.7 s and
a frequency window of 2.6 MHz centered around 19.1 MHz. The emission
is seen to consist of several narrow bands of emission at frequencies
near 20 MHz as well as S-bursts at frequencies between 18 and 20 MHz.
We note that the horizontal structures seen in the figure are
RFI. Analysis of additional sections of the Io-D emission show
structures such as tilted-V events and other narrow-band structures.

More observations are currently being conducted in an effort to
  further characterize the fine structure of the Io-D events which can
  be useful in understanding the electron acceleration method and the
  source generation mechanism.

\section{Summary}

The Long Wavelength Array (LWA1) instrument is shown to be an
excellent instrument for studies of Jupiter decametric radio emission
(DAM). The large bandwidth, fine spectral and temporal resolution, and
full Stokes capabilities of the LWA1 instrument allow for a
comprehensive analysis of Jupiter's broad and fine DAM structures. The
location and low RFI environment are suitable to measure Jupiter's
emissions down to about 11 MHz, well below most other Earth-based
radio telescopes. This overview paper of targeted Io-related Jovian
DAM emissions not only highlights the capabilities of the LWA1
instrument, but also provides some new data and insight into the
nature of Jupiter's decametric emissions.  The Jupiter observations
from Dec-2011 to Mar-2013 highlighted in this paper show good Io-A,
Io-B, Io-C, and Io-D structures, and the broad and fine detail in
these data are further enhanced by the excellent polarization
capabilities of the LWA1 instrument. Data reduction and analysis show
some characteristic Io-related spectral signatures but also show some
fine structures and details not seen before. A summary of the findings
follows.

The Io-A observations show characteristic vertex-late arc structures
as well as some narrow band emission events at frequencies below 19
MHz (see Figure~\ref{fig:262}). The remarkable observations are the
Io-C emissions that occur simultaneously with the Io-A emissions. The
onset of the LHC Io-C emissions at 11 MHz was measured to be
\mbox{CML\,{\sc iii}} 230$^\circ$; this has not been clearly measured
before and allows the extent of the Io-C source to be significantly
expanded. The CML range of the Io-C source can tentatively be defined
as \mbox{CML\,{\sc iii}} 230$^\circ$ $-$ 60$^\circ$ \citep[previous
  range was 280$^\circ$ $-$ 60$^\circ$]{carr83}.  Further observations
are warranted, but this can give clues to the nature and geometry of
the southern hemisphere emission regions and be of value to new
decameter emission models.  In addition, the RHC component emissions
from Io-C are seen in excellent detail; these observations, along with
modulation lane measurements, give additional evidence that may
suggest that both RHC and LHC emissions are coming from the same
hemisphere source regions.

Observations of Io-B and Io-D emissions are clearly seen with the dual
polarization LWA1 measurements. The Io-B data show characteristic arc
and nested band structures, normal envelope and tail structures due to
source geometry and beaming, and characteristic S-bursts that occur
early in the Io-B events. The Io-D events are easily seen and analyzed
with the LWA1 instrument. The Io-D events show characteristic LHC
polarization and one event (Figure~\ref{fig:997_full}) shows narrow band
structure. 

The LWA1 instrument shows excellent fine structure in Jupiter's
emission. Modulation lanes are clearly seen in both polarizations and
drift rate measurements are consistent with earlier results. However,
modulation lane measurements from one Io-A/Io-C event where both
polarizations are observed simultaneously (Figure~\ref{fig:mod_lanes})
might provide good constraints on the emission theory and growth
mechanisms. More detailed analysis of additional Io-A/Io-C events are
underway to further investigate the modulation lane details. Faraday
fringes are also clearly seen in the LWA1 data and provide clues about
the local conditions at the emission source, in the Io plasma torus,
and the Earth's ionosphere.

Finally, and perhaps most interestingly, we have made excellent
spectral and temporal measurements of S-bursts, N-events, and S-N
events. First we have clearly detected S-bursts in the Io-D source, a
measurement that is rare and will be useful to help understand the
mechanism of S-burst emission. Drift rates of -12 MHz/s were measured
for the Io-D region and are consistent with known drift rates from
other Io-related sources. Several N-events are seen in our data in
Io-A, Io-B, and Io-D events, but most interesting are the S-N events
seen in an Io-B observation
(Figures~\ref{fig:IoB_SN_wide}-\ref{fig:IoB_SN_zoom}). These events
display complex interaction where S-bursts appear to quench N-events,
and N-events may trigger more S-bursts. These effects demonstrate that
the emission mechanisms for the S-N events are similar and/or
interacting, and that they are generated in the same spatial
location. As mentioned previously, more observations of Io-D are being
made in anticipation of further study of the S-bursts measured with
the LWA1 instrument.


%

\begin{acknowledgments}
We thank the referees for a careful reading of the manuscript and
helpful comments. We also wish to thank the staff of the Long
Wavelength Array. Basic research in radio astronomy at the Naval
Research Laboratory is supported by 6.1 Base funding. Construction of
LWA1 was supported by the Office of Naval Research under Contract
N00014-07-C- 0147. Support for operations and continuing development
of LWA1 is provided by the National Science Foundation under grants
AST-1139963 and AST-1139974 of the University Radio Observatories
program. CAH acknowledges support from the Tennessee Space Grant
Consortium. This research has been supported in part by JSPS KAKENHI
Grant Number 25400480. JS (a student at Montgomery Blair High School) 
was supported by the Science and Engineering
Apprenticeship Program (SEAP) at the Naval Research Laboratory. GBT
acknowledges partial support for this work from NSF grant
AST-1212162. Data used for this publication are available through
request to the authors.
\end{acknowledgments}

%

%
%

\end{article}

\end{document}